\documentclass[aps,amsmath,showpacs]{revtex4}
\usepackage{bm}
\usepackage{epsfig}
\begin{document}

\title{Frequency and damping of hydrodynamic modes in a trapped Bose-condensed gas}

\author{Tetsuro Nikuni}
\affiliation{Department of Physics, Faculty of Science, 
Tokyo University of Science, 1-3 Kagurazaka, Shinjuku-ku,
Tokyo, Japan 162-9601}

\author{Allan Griffin}
\affiliation{Department of Physics, University of Toronto, Toronto, Ontario,
Canada M5S 1A7}

\date{\today}
\begin{abstract}
Recently it was shown that the Landau-Khalatnikov two-fluid hydrodynamics describes 
the collision-dominated region of a trapped Bose condensate interacting with a
thermal cloud.
We use these equations to discuss the low frequency hydrodynamic collective modes
in a trapped Bose gas at finite temperatures.
We derive a variational expressions based on these equations for both the frequency and
damping of collective modes.
A new feature is our use of frequency-dependent transport coefficients,
which produce a natural cutoff by eliminating the collisionless low-density tail of the
thermal cloud.
Above the superfluid transition, our expression for the damping in trapped inhomogeneous
gases is analogous to the result
first obtained by Landau and Lifshitz for uniform classical fluids.
We also use the moment method to discuss the crossover from the collisionless to
the hydrodynamic region.
Recent data for the monopole-quadrupole mode in the hydrodynamic region of a
trapped gas of metastable $^4$He is discussed.
We also present calculations for the damping of the analogous $m=0$ monopole-quadrupole
condensate mode in the superfluid phase.

\end{abstract}
\maketitle

\section{introduction}
In recent papers, Zaremba and the authors have derived a closed set of the 
two-fluid hydrodynamic equation of a trapped Bose-condensed gas starting from a 
simplified microscopic model describing the coupled dynamics of the condensate and 
noncondensate atoms~\cite{ZGN,NZG,ZNG,CJP,LK}.
These equations can be written in the Landau-Khalatnikov (LK) form,
well known in the study of superfluid $^4$He \cite{Khalatnikov,Wilks}.
These simplified hydrodynamic equations include dissipative terms associated
with the shear viscosity, the thermal conductivity, and the four
second-viscosity coefficients.
Explicit formulas for these transport coefficients were obtained in Ref.~\cite{LK} and
used to define three characteristic transport relaxation times \cite{relax}.
These define the crossover between the collisionless and hydrodynamic regions.
Detailed calculations of these transport relaxation times in a trapped Bose
gas \cite{relax} shows that the collisions between the condensate and
noncondensate enhances the transport relaxation rates significantly
in the MIT data \cite{MIT}, so that one is in the hydrodynamic region below $T_{\rm BEC}$.
We also note that the recent Bose condensate observed in metastable He$^*$ 
\cite{ENS1,ENS2,ENS3}
appears to be well within the collision-dominated hydrodynamic region, even above the
Bose-Einstein condensation temperature $T_{\rm BEC}$.
This is because of the relatively large density of He$^*$ atoms and their large $s$-wave 
scattering length.

In the present paper, we derive a general expression for the frequency and damping of
hydrodynamic collective modes in a trapped Bose-condensed gas at finite temperatures,
starting from the two-fluid hydrodynamic equations derived in Ref.~\cite{LK}.
These two-fluid equations are briefly reviewed in Section \ref{sec:two-fluid}, and reformulated as a 
closed set of equations for the condensate and noncondensate velocity fields.
In Section \ref{sec:frequency}, we derive a variational expression for undamped normal-mode
frequencies in the Landau limit ($\omega\tau\ll 1$), extending an approach first developed
in Ref.~\cite{ZNG}.
In Section \ref{sec:damping}, we obtain a general expression for the damping, which only depends on
knowing the undamped normal-mode solutions.
This kind of expression is very convenient in working out the damping of hydrodynamic modes
in trapped Bose gases, as first pointed out by Kavoulakis et al.~\cite{KPS}. 
As an illustration, in Section \ref{sec:aboveTc}  we give a detailed discussion of the
the $m=0$ monopole-quadrupole collective mode above $T_{\rm BEC}$ studied in the recent
experiments \cite{ENS3}.
In section \ref{sec:belowTc}, we also calculate the damping of the $m=0$ hydrodynamic mode 
in the superfluid phase.

Appendix A gives some details of the damping calculations based on the use of 
frequency-dependent transport coefficients.
The moment method for a degenerate normal Bose gas is reviewed in Appendix B.


\section{Two-fluid hydrodynamics of a trapped Bose gas: A review}
\label{sec:two-fluid}
In this paper, we consider a Bose-condensed gas confined in an external anisotropic
harmonic trap potential
\begin{equation}
U_{\rm ext}({\bf r})=\frac{m}{2}(\omega_x^2x^2+\omega_y^2y^2+\omega_z^2z^2),
\label{eq:trap}
\end{equation}
Our starting point is the two-fluid hydrodynamics derived from the finite-temperature
kinetic theory by Zaremba, Nikuni, and Griffin (ZNG) \cite{ZNG}.
In the ZNG theory, the coupled dynamics of the condensate and noncondensate~\cite{ZNG} is
described by the generalized Gross-Pitaevskii (GP) equation for the
condensate wavefunction $\Phi({\bf r},t)$
\begin{equation}
i\hbar\frac{\partial\Phi({\bf r},t)}{\partial t}=\left[-\frac{\hbar^2\nabla^2}{2m}
+U_{\rm ext}({\bf r})+gn_c({\bf r},t)+2g\tilde n({\bf r},t)-iR({\bf r},t)\right]
\Phi({\bf r},t), 
\label{eq:GP}
\end{equation}
and the semi-classical kinetic equation for the noncondensate distribution
function $f({\bf r},{\bf p},t)$
\begin{equation}
\frac{\partial f({\bf r},{\bf p},t)}{\partial t}+\frac{{\bf p}}{m}\cdot\bm{\nabla}_{\bf r} 
f({\bf r},{\bf p},t)-\bm{\nabla}U({\bf r},t)\cdot\bm{\nabla}_{\bf p} f({\bf r},{\bf p},t)
=C_{12}[f,\Phi]+C_{22}[f].
\label{eq:QK}
\end{equation}
Here $n_c({\bf r},t)=|\Phi({\bf r},t)|^2$ is the condensate density,
and $\tilde n({\bf r},t)$ is the noncondensate density,
\begin{equation}
\tilde n({\bf r},t)=\int \frac{d{\bf p}}{(2\pi\hbar)^3} f({\bf r},{\bf p},t),
\label{eq:ntilde}
\end{equation}
and $U({\bf r},t)=U_{\rm ext}({\bf r})+2g[n_c({\bf r},t)+\tilde n({\bf r},t)]$
is the time-dependent effective potential acting on the noncondensate,
including the Hartree-Fock (HF) mean field.
As usual, we approximate the interaction in the $s$-wave scattering approximation
$g=4\pi\hbar^2a/m$.
The dissipative term $R({\bf r},t)$ in the generalized GP equation (\ref{eq:GP})
is due to the collisional exchange of atoms in the condensate and noncondensate.
This is related to the $C_{12}$ collision integral in (\ref{eq:QK}), namely
\begin{equation}
R({\bf r},t)=\frac{\hbar\Gamma_{12}({\bf r},t) } {2n_c({\bf r},t)},~~~
\Gamma_{12}({\bf r},t)=\int\frac{d{\bf p}}{(2\pi\hbar)^3}C_{12}[f({\bf r},{\bf p},t),
\Phi({\bf r},t)].
\label{R_term}
\end{equation}
Explicit expressions for the two collision integrals ($C_{22}$ and $C_{12}$) in the kinetic
equation (\ref{eq:QK}) can be found in Ref.~\cite{ZNG}.

The GP equation (5) can be written in the hydrodynamic form 
in terms of the amplitude and phase of 
$\Phi({\bf r},t)=\sqrt{n_c({\bf r},t)}e^{i\theta({\bf r},t)}$, which leads to
\begin{subequations}
\begin{eqnarray}
\frac{\partial n_c}{\partial t} + \bm{\nabla}\cdot(n_c{\bf v}_c)&=& 
-\Gamma_{12}[f,\Phi]\,,  
\label{eq_nc}
\\
m\left(\frac{\partial}{\partial t}+{\bf v}_c\cdot 
\bm{\nabla}\right) {\bf v}_c&=&-\bm{\nabla}\mu_c \ ,
\label{eq_vc}
\end{eqnarray}
\label{hydro-C}
\end{subequations}

\noindent
where the superfluid velocity is ${\bf v}_c\equiv\hbar \bm{\nabla}\theta({\bf r},t)/m$ 
and the condensate chemical potential is given by
\begin{equation}
\mu_c({\bf r}, t) =-\frac{\hbar^2\nabla^2\sqrt{n_c({\bf r},t)} }{2m\sqrt{n_c({\bf r},t)}}
+U_{\rm ext}({\bf r})+
gn_c({\bf r}, t)+2g\tilde{n}({\bf r}, t)\, .
\label{mu_c}
\end{equation}
One sees that $\Gamma_{12}$ in Eq.~(\ref{eq_nc}) plays the role of a ``source function" in the 
continuity equation for the condensate, arising from the fact that $C_{12}$ collisions
do not conserve the number of condensate atoms \cite{ZNG}.

Hydrodynamic equations for the noncondensate can be derived by
following the standard procedure first developed in the kinetic theory of classical gases.
We take moments of the kinetic equation (\ref{eq:QK}) with respect to $1,{\bf p}$ and $p^2$
to derive the most general form of ``hydrodynamic-type equations" for the
non-condensate. These moment equations take the form ($\mu$ and $\nu$ 
are Cartesian components):
\begin{subequations}
\begin{eqnarray}
&&\frac{\partial{\tilde n}}{\partial t}+\bm{\nabla}\cdot 
(\tilde{n}{\bf v}_n) = \Gamma_{12}[f]\,, 
\label{hydro_generala}
\\
&&m{\tilde n}\left(\frac{\partial}{\partial t}+{\bf v}_n\cdot 
\bm{\nabla}\right) v_{n\mu}=-\frac{\partial P_{\mu\nu}}{\partial x_\nu}
-{\tilde n}\frac{\partial U}{\partial x_\mu}
-m(v_{n\mu}-v_{c\mu})\Gamma_{12}[f]\,, 
\label{hydro_generalb}
\\
&&\frac{\partial\tilde\epsilon}{\partial t} +
\nabla\cdot(\tilde\epsilon{\bf v}_n) = -\bm{\nabla}\cdot{\bf Q}
-D_{\mu\nu} P_{\mu\nu}+ \left[\frac{1}{2}m({\bf v}_n-{\bf v}_c)^2
+\mu_c-U\right]\Gamma_{12}[f]. 
\label{hydro_generalc}
\end{eqnarray}
\label{hydro_general}
\end{subequations}

\noindent
Here and elsewhere, repeated Greek subscripts are summed.
The noncondensate density was defined earlier in Eq.~(\ref{eq:ntilde}), 
while the noncondensate local velocity ${\bf v}_n({\bf r},t)$ is defined by
\begin{equation}
{\tilde n}({\bf r},t){\bf v}_n({\bf r}, t)\equiv\int\frac{d{\bf p}}
{(2\pi\hbar)^3} \frac{{\bf p}}{m} f({\bf r, p}, t)\,.
\label{vn}
\end{equation}
In addition, we have introduced the following quantities:
\begin{subequations}
\begin{eqnarray}
P_{\mu\nu}({\bf r}, t)&\equiv& m
\int\frac{d{\bf p}}{(2\pi\hbar)^3}\left(\frac{p_\mu}{m} - v_{n\mu}\right)
\left(\frac{p_\nu}{m} - v_{n\nu}\right)
f({\bf r, p}, t), \label{eq35a} \\
{\bf Q}({\bf r}, t)&\equiv& \int\frac{d{\bf p}}{(2\pi\hbar)^3} \frac{1}{2m} 
({\bf p}-m{\bf v}_n)^2\left(\frac{{\bf p}}{m}-{\bf v}_n\right)
f({\bf r, p}, t),\label{eq35b}  \\
\tilde \epsilon({\bf r}, t) &\equiv&\int\frac{d{\bf p}}{(2\pi\hbar)^3}
\frac{1}{2m} ({\bf p}-m{\bf v}_n)^2
f({\bf r},  {\bf p}, t) \,. 
\end{eqnarray}
\label{pqe}
\end{subequations}
Finally, the symmetric rate-of-strain tensor appearing in Eq.~(\ref{hydro_generalc}) 
is defined as
\begin{equation}
D_{\mu\nu}({\bf r}, t) \equiv \frac{1}{2} 
\left(\frac{\partial v_{n \mu}}{\partial x_\nu} + \frac{\partial v_{n \nu}}{\partial x_\mu}
\right).
\label{Dmunu}
\end{equation}
The ``hydrodynamic'' equations (\ref{hydro_general})-(\ref{Dmunu}) are exact, but
not closed as they stand.

We next apply the Chapman-Enskog procedure to obtain a closed set of
hydrodynamic equations.
This procedure yields~\cite{LK}
\begin{equation}
P_{\mu\nu}=\delta_{\mu\nu}\tilde P
-2\eta\left(D_{\mu\nu}-\frac{1}{3}{\rm Tr}D \delta_{\mu\nu}\right), 
\end{equation}
\begin{equation}
{\bf Q}=-\kappa\bm{\nabla}T,
\end{equation}
where $\tilde P$ is the kinetic pressure and $T$ is the temperature.
The above formulas involve the position-dependent shear viscosity $\eta$ and the thermal
conductivity $\kappa$.
These local position-dependent transport coefficients will be discussed in more detail 
below.

In order to study small amplitude oscillations, we linearize the hydrodynamic
equations around static equilibrium as
$n_c=n_{c0}+\delta n_c,{\bf v}_c=\delta {\bf v}_c,\tilde n=\tilde n_0+\delta\tilde n,
{\bf v}_n=\delta {\bf v}_n, \tilde P=\tilde P_0+\delta P$,
where the subscript 0 denotes static equilibrium.
The equilibrium condensate density profile is determine by (within the Thomas-Fermi 
approximation, which neglects the quantum pressure or kinetic energy term in the GP equation)
\begin{equation}
n_{c0}({\bf r})=\frac{1}{g}[\mu_{c0}-U_{\rm ext}({\bf r})]-2\tilde n_0({\bf r}),
\label{nc0}
\end{equation}
while the equilibrium noncondensate distribution function $f_0$ is given by
\begin{equation}
f_0({\bf r},{\bf p})=\frac{1}{e^{\beta_0[\frac{p^2}{2m}+U_0({\bf r})-\mu_{c0}]}-1}.
\label{eq:f0}
\end{equation}
The local density $\tilde n_0({\bf r})$ and the local kinetic pressure
$\tilde P_0({\bf r})$ of the noncondensate atoms are given from Eq.~(\ref{eq:f0}) as
\begin{equation}
\tilde n_0({\bf r})=\frac{1}{\Lambda^3}g_{3/2}(z_0),
\label{ntilde0}
\end{equation}
\begin{equation}
\tilde P_0({\bf r})=\frac{k_{\rm B}T_0}{\Lambda^3}g_{5/2}(z_0), 
\label{ptilde0}
\end{equation}
where $z_0({\bf r})=e^{[\mu_{c0}-U_0({\bf r})]/k_{\rm B}T_0}$ is the local
equilibrium fugacity,
$\Lambda =(2\pi\hbar^2/mk_{\rm B}T_0)^{1/2}$ 
is the thermal de Broglie wavelength,
and $g_n(z)=\sum_{l=1}^{\infty}z^l/l^n$.

The linearized hydrodynamic equations for the condensate are given by
\begin{subequations}
\begin{eqnarray}
\frac{\partial\delta n_c}{\partial t}&=&-\bm{\nabla}\cdot(n_{c0}\delta{\bf v}_c)
-\delta\Gamma_{12}, \label{lineq_nc}\\
m\frac{\partial\delta{\bf v}_c}{\partial t}&=&-g\bm{\nabla}\delta
(\delta n_c+2\delta \tilde n), \label{lineq_vc}
\end{eqnarray}
\label{eq_C}
\end{subequations}
while the linearized hydrodynamic equations for the noncondensate atoms
are given by
\begin{subequations}
\begin{eqnarray}
&&\frac{\partial \delta \tilde n}{\partial t}+\bm{\nabla}
 \cdot(\tilde n_0\delta{\bf v}_n)=\delta\Gamma_{12}, \label{eq_ntilde}\\
&&m\tilde n_0\frac{\partial \delta v_{n\mu}}{\partial t}
=-\frac{\partial \delta \tilde P}{\partial x_{\mu}} 
- \delta \tilde n \frac{\partial U_0}{\partial x_{\mu}} 
-2g\tilde n_0\frac{\partial \delta n}{\partial x_{\mu}} +
\frac{\partial}{\partial x_{\nu}}\left\{2\eta\left[D_{\mu\nu}-\frac{1}{3}
({\rm Tr}D)\delta_{\mu\nu} \right]\right\}, \label{eq_vn}\\
&&\frac{\partial \delta \tilde P}{\partial t}=
-\frac{5}{3}\bm{\nabla}\cdot(\tilde P_0\delta {\bf v}_n)
+\frac{2}{3}\delta{\bf v}_n\cdot\bm{\nabla} \tilde P_0
+(\mu_{c0}-U_0)\delta\Gamma_{12}
+\frac{2}{3}\bm{\nabla}\cdot(\kappa\bm{\nabla} 
\delta T). \label{eq_ptilde}
\end{eqnarray}
\label{eq_NC}
\end{subequations}
The above equations involve the fluctuations of the source function
$\delta\Gamma_{12}$ and the temperature $\delta T$.
These can also be written in terms of the condensate and noncondensate
velocity fields \cite{LK}. One finds
$\delta\Gamma_{12}=\delta \Gamma_{12}^{(1)}+\delta\Gamma_{12}^{(2)}$, where
\begin{eqnarray}
\delta \Gamma_{12}^{(1)}&=&\sigma_H\left\{
\bm{\nabla}\cdot[n_{c0}(\delta{\bf v}_c-\delta{\bf v}_n)]
+\frac{1}{3}n_{c0}\bm{\nabla}\cdot\delta{\bf v}_n\right\},
\label{G12_1} \\
\delta\Gamma_{12}^{(2)}&=&-\tau_{\mu}\frac{\partial}{\partial t}\delta
\Gamma_{12}^{(1)}-\frac{2\sigma_H\sigma_1}{3g\tilde n_0}\bm{\nabla}
\cdot(\kappa\bm{\nabla}\delta T), \\
\frac{\partial\delta T}{\partial t}&=&-\frac{2}{3}T_0\bm{\nabla}\cdot\delta{\bf v}_n
+\frac{2T_0}{3\tilde n_0}\sigma_1\delta\Gamma_{12}^{(1)}.
\end{eqnarray}
Here $\sigma_H$ and $\sigma_1$ only involve the static local thermodynamic
functions of the gas, and are defined in Eqs.~(25) and (57) of Ref.~\cite{LK}.
In Eq.~(\ref{G12_1}), $\tau_{\mu}$ is a new relaxation time describing how fast the condensate
and noncondensate atoms reach diffusive local equilibrium with each other.
More explicitly, it is given by
\begin{equation}
\frac{1}{\tau_{\mu}}=\left(\frac{gn_{c0}}{k_{\rm B}T}\right)
\frac{1}{\tau_{12}\sigma_H},
\end{equation}
where $\tau_{12}$ is a collision time of the condensate atoms with the
noncondensate atoms as defined in Ref.~\cite{LK}.

The above two-fluid hydrodynamic equations have dissipative terms involving the
shear viscosity $\eta$ and the thermal conductivity $\kappa$.
These transport coefficients are given by the following expressions \cite{LK,relax}
\begin{equation}
\kappa=\frac{5}{2}\tau_{\kappa}
\frac{\tilde n_0 k_{\rm B}^2 T_0}{m}
\left\{ 
\frac{7g_{7/2}(z_0)}{2g_{3/2}(z_0)}-
\frac{5}{2}
\left[ \frac{g_{5/2}(z_0)} {g_{3/2}(z_0)} \right]^2
\right\},
\label{eq:kappa}
\end{equation}
\begin{equation}
\eta=\tau_{\eta}\tilde n_0k_{\rm B}T_0\left[\frac{g_{5/2}(z_0)}{g_{3/2}(z_0)}\right],
\label{eq:eta}
\end{equation}
where $z_0$ is the local fugacity defined earlier.
These expressions involve the characteristic transport relaxation times
$\tau_{\kappa}$ and $\tau_{\eta}$, which are defined in Refs.~\cite{LK,relax}.
The three relaxation times $\tau_{\kappa},\tau_{\eta}$ and $\tau_{\mu}$
characterize how fast the two-component system reaches local equilibrium,
and thus they define the crossover frequency between the collision-dominated hydrodynamic
region and the so-called collisionless region dominated by mean fields.
We note that the two-fluid equations (\ref{eq_C}) and (\ref{eq_NC}) can also be
rewritten in the LK form \cite{LK,Khalatnikov}, which involve the four second viscosity
coefficients $\zeta_1,\zeta_2,\zeta_3$, and $\zeta_4$.
As shown in Ref.~\cite{LK}, they are all related to the relaxation time $\tau_{\mu}$
as follows:
\begin{eqnarray}
\zeta_1&=&\zeta_4=\frac{gn_{c0}}{3m}\sigma_H\tau_{\mu}, \cr
\zeta_2&=&\frac{gn_{c0}^2}{9}\sigma_H\tau_\mu,~~
\zeta_3=\frac{g}{m^2}\sigma_H\tau_{\mu}.
\end{eqnarray}

To derive a closed set of equations for the velocity fields ${\bf v}_c$
and ${\bf v}_n$, we take time derivatives of (\ref{eq_vc}) and (\ref{eq_vn}).
We obtain
\begin{equation}
m\frac{\partial^2 \delta{\bf v}_c}{\partial t^2}
=g\bm{\nabla}[\bm{\nabla}\cdot(n_{c0}\delta{\bf v}_c)
+2\bm{\nabla}\cdot(\tilde n_0\delta{\bf v}_n)]-g\bm{\nabla}\delta\Gamma_{12},
\label{eq:vc}
\end{equation}
and
\begin{eqnarray}
m\tilde n_0\frac{\partial^2 \delta v_{n\mu} }{\partial t^2}
&=&\frac{5}{3}\frac{\partial}{\partial x_{\mu}}[\bm{\nabla}\cdot
(\tilde P_0{\bf v}_n)]-\frac{2}{3}\frac{\partial}{\partial x_{\mu}}
(\delta{\bf v}_n\cdot\bm{\nabla}\tilde P_0)
+\bm{\nabla}\cdot(\tilde n_0\delta{\bf v}_n)
\frac{\partial U_0}{\partial x_{\mu}} \cr
&&+2g\tilde n_0\frac{\partial}{\partial x_{\mu}}
[\bm{\nabla}\cdot(n_{c0}\delta{\bf v}_c)
+\bm{\nabla}\cdot(\tilde n_0\delta{\bf v}_n)] \cr
&&-\frac{1}{3}\frac{\partial U_0}{\partial x_{\mu}}\delta\Gamma_{12}
+\frac{2}{3}gn_{c0}\frac{\partial\delta\Gamma_{12}}{\partial x_{\mu}} \cr
&&+\frac{\partial}{\partial x_{\nu}}\left\{2\eta\left[\frac{\partial}{\partial t}
D_{\mu\nu}-\frac{1}{3}\left({\rm Tr}\frac{\partial D}{\partial t}\right)
\delta_{\mu\nu}\right]\right\}-\frac{2}{3}\frac{\partial}{\partial x_{\mu}}
\bm{\nabla}\cdot(\kappa\bm{\nabla}\delta T). \label{eq:vn}
\end{eqnarray}
We then look for normal-mode solutions of (\ref{eq:vc}) and (\ref{eq:vn}) of the
form
\begin{equation}
{\bf v}_n({\bf r},t)={\bf u}_n({\bf r})e^{-i\omega t}, \ \ 
{\bf v}_c({\bf r},t)={\bf u}_c({\bf r})e^{-i\omega t}.
\end{equation}
In this case, the coupled equations (\ref{eq:vc}) and (\ref{eq:vn}) reduce to
\begin{equation}
m\omega^2{\bf u}_c=-g\bm{\nabla}[\bm{\nabla}\cdot(n_{c0}{\bf u}_c)]
-2g\bm{\nabla}[\bm{\nabla}\cdot(\tilde n_0{\bf u}_n)]+g\bm{\nabla}\delta
\Gamma_{12,\omega}[{\bf u}_n,{\bf u}_c],
\label{eq:uc}
\end{equation}
\begin{eqnarray}
m\tilde n_0\omega^2 u_{n\mu}
&=& 
-\frac{5}{3}\frac{\partial}{\partial x_{\mu}}[\bm{\nabla}\cdot
(\tilde P_0{\bf u}_n)]
+\frac{2}{3}\frac{\partial}{\partial x_{\mu}}
({\bf u}_n\cdot\bm{\nabla}\tilde P_0)
-\bm{\nabla}\cdot(\tilde n_0{\bf u}_n)
\frac{\partial U_0}{\partial x_{\mu}} \cr
&&-2g\tilde n_0\frac{\partial}{\partial x_{\mu}}
[\bm{\nabla}\cdot(n_{c0}{\bf u}_c)
+\bm{\nabla}\cdot(\tilde n_0{\bf u}_n)] \cr
&&+\frac{1}{3}\frac{\partial U_0}{\partial x_{\mu}}
\delta\Gamma_{12,\omega}[{\bf u}_n,{\bf u}_c] 
-\frac{2}{3}gn_{c0}\frac{\partial\delta\Gamma_{12,\omega}[{\bf u}_n,{\bf u}_c]}
{\partial x_{\mu}} \cr
&&+i\omega\frac{\partial}{\partial x_{\nu}}\left\{2\eta\left[
u_{\mu\nu}-\frac{1}{3}\left(\bm{\nabla}\cdot{\bf u}_n\right)
\delta_{\mu\nu}\right]\right\}
+\frac{2}{3}\frac{\partial}{\partial x_{\mu}}
\bm{\nabla}\cdot(\kappa\bm{\nabla}\delta T_{\omega}[{\bf u}_n,{\bf u}_c]).
\label{eq:un}
\end{eqnarray}
The symmetric tensor $u_{\mu\nu}$ is defined by
\begin{equation}
u_{\mu\nu}\equiv\frac{1}{2}\left(\frac{\partial u_{n\nu}}{\partial x_{\mu}}
+\frac{\partial u_{n\mu}}{\partial x_{\nu}}\right).
\end{equation}

The source function $\delta\Gamma_{12}$ appearing in the above equations can be expressed 
in terms of of ${\bf u}_n$ and ${\bf u}_c$ as
$\delta\Gamma_{12}=\delta\Gamma_{12,\omega}[{\bf u}_n,{\bf u}_c]e^{-i\omega t}$, where
\begin{equation}
\delta\Gamma_{12,\omega}[{\bf u}_n,{\bf u}_c]=
\delta\Gamma_{12}^{(1)}[{\bf u}_n,{\bf u}_c]+
\delta\Gamma_{12,\omega}^{(2)}[{\bf u}_n,{\bf u}_c],
\end{equation}
with
\begin{equation}
\delta\Gamma_{12}^{(1)}[{\bf u}_n,{\bf u}_c]
=\sigma_H\left\{\bm{\nabla}\cdot[n_{c0}({\bf u}_c-{\bf u}_n)]+\frac{1}{3}n_{c0}
\bm{\nabla}\cdot{\bf u}_n\right\}.
\label{gamma121}
\end{equation}
\begin{equation}
\delta\Gamma_{12,\omega}^{(2)}=i\omega\tau_{\mu}\delta\Gamma_{12}^{(1)}[{\bf u}_n,{\bf u}_c]
-\frac{2\sigma_H\sigma_1}{3g\tilde n_0}\bm{\nabla}\cdot(\kappa\delta T_{\omega}
[{\bf u}_n,{\bf u}_c]).
\end{equation}
Similarly, the temperature fluctuation is given by
$\delta T=i\delta T_{\omega}[{\bf u}_n,{\bf u}_c]$, where
\begin{equation}
\delta T_{\omega}[{\bf u}_n,{\bf u}_c]=\frac{1}{\omega}\left[
-\frac{2}{3}T_0(\bm{\nabla}\cdot{\bf u}_n)+\frac{2T_0}{3\tilde n_0}
\sigma_1\delta\Gamma_{12}^{(1)}[{\bf u}_n,{\bf u}_c]\right].
\label{tomega}
\end{equation}
Using these results in Eqs.~(\ref{eq:uc}) and (\ref{eq:un}), we see that we have
obtained a {\it closed} set of equations for both local velocity components
${\bf u}_n$ and ${\bf u}_c$.

\section{Undamped normal mode frequency}
\label{sec:frequency}
We first consider the undamped normal-mode solutions of our hydrodynamic equations,
neglecting  all hydrodynamic dissipation.
Formally this means that we take $\eta,\kappa,\tau_{\mu}\to0$ in the two-fluid
hydrodynamic equations.
As discussed in Refs.~\cite{ZNG,LK},
this limit corresponds to the Landau two-fluid hydrodynamics without
dissipation.
In this limit, the coupled equations for ${\bf u}_n$ and ${\bf u}_c$
simplify to
\begin{eqnarray}
m\omega^2{\bf u}_c&=&-g\bm{\nabla}[\bm{\nabla}\cdot(n_{c0}{\bf u}_c)]
-2g\bm{\nabla}[\bm{\nabla}\cdot(\tilde n_0{\bf u}_n)]+g\bm{\nabla}
\delta\Gamma_{12,\omega}^{(1)}[{\bf u}_n,{\bf u}_c], \label{eq:uc2}
\\
m\omega^2{\bf u}_n&=&-\frac{5}{3}\bm{\nabla}[\bm{\nabla}\cdot(\tilde P_0{\bf u}_n)]
+\frac{2}{3}\bm{\nabla}[{\bf u}_n\cdot\bm{\nabla}\tilde P_0]-\frac{2}{3}
\bm{\nabla}\left(gn_{c0}\delta\Gamma_{12,\omega}^{(1)}[{\bf u}_n,{\bf u}_c]\right) \cr
&&-\delta \Gamma_{12}^{(1)}[{\bf u}_n,{\bf u}_c]\bm{\nabla}U_0
-\bm{\nabla}\cdot(\tilde n_0{\bf u}_n)\bm{\nabla}U_0
-2g\tilde n_0\bm{\nabla}[\bm{\nabla}\cdot(\tilde n_0{\bf u}_n)
+\bm{\nabla}\cdot(n_{c0}{\bf u}_c)],
\label{eq:un2}
\end{eqnarray}
In general, solutions of these coupled hydrodynamic equations are very complicated.
However, one can reformulate this problem so that the solutions are given in terms
of a variational functional.
The present discussion closely follows the variational analysis developed in Ref.~\cite{ZNG} for
two-fluid hydrodynamic equations, introduced in Ref.~\cite{ZGN}, which omitted the contribution
of the source term from $C_{12}$ collisions, i.e., $\delta\Gamma_{12}=0$.
However, as Ref.~\cite{ZNG} showed, $\delta\Gamma_{12}$ plays a crucial role in obtaining
the correct Landau two-fluid hydrodynamic limit.
The same formalism was also used in Ref.~\cite{cigar} to calculated the hydrodynamic 
mode frequencies for a trapped Bose gas above $T_{\rm BEC}$.

Introducing the six-component local velocity vector 
\begin{equation}
{\bf u}=
\left(
\begin{matrix}
{\bf u}_n \cr
{\bf u}_c
\end{matrix}
\right),
\end{equation}
we combine the coupled equations for ${\bf u}_n$ and ${\bf u}_c$ in (\ref{eq:un2}) and (\ref{eq:uc2})
into a matrix equation
\begin{equation}
\bm{L}{\bf u}=\omega^2\bm{D}{\bf u}.
\label{matrix_eq}
\end{equation}
The $6\times 6$ matrix $\bm{L}$ has the block structure,
\begin{equation}
{\bm L} = 
\left(
\begin{matrix}
\hat L_{11}&\hat L_{12}\cr
\hat L_{21}&\hat L_{22}
\end{matrix}
\right),
\label{matrixL}
\end{equation}
with the $3\times 3$ matrix elements being defined as
\begin{subequations}
\begin{eqnarray}
(\hat L_{11}{\bf u}_n)_{\mu} &=& 
-\frac{5}{3}\frac{\partial}{\partial x_{\mu}}[\bm{\nabla}\cdot
(\tilde P_0{\bf u}_n)]
+\frac{2}{3}\frac{\partial}{\partial x_{\mu}}
({\bf u}_n\cdot\bm{\nabla}\tilde P_0)  \cr
&&-\bm{\nabla}\cdot(\tilde n_0{\bf u})
\frac{\partial U_0}{\partial x_{\mu}}
-2g\tilde n_0\frac{\partial}{\partial x_{\mu}}
[\bm{\nabla}\cdot(\tilde{n}_{0}{\bf u}_n)] \cr
&&-\frac{\partial}{\partial x_{\mu}}
\left(\frac{2}{3}gn_{c0}\delta\Gamma_{12}^{(1)}[{\bf u}_n,0]\right) 
+\delta\Gamma_{12}^{(1)}[{\bf u}_n,0]\frac{\partial U_0}{\partial x_{\mu}},
\label{L11} \\
(\hat L_{12}{\bf u}_c)_{\mu} &=& -2g\tilde n_0 \frac{\partial}{\partial x_{\mu}}
\bm{\nabla}\cdot(n_{c0}{\bf u}_c)
-\frac{\partial}{\partial x_{\mu}}
\left(\frac{2}{3}gn_{c0}\delta\Gamma_{12}^{(1)}[0,{\bf u}_c]\right) \cr 
&&+\delta\Gamma_{12}^{(1)}[0,{\bf u}_c]\frac{\partial U_0}{\partial x_{\mu} },  
\label{L12} \\
(\hat L_{21}{\bf u}_n)_{\mu} &=& -2g n_{c0} \frac{\partial}{\partial x_{\mu}}
\bm{\nabla}\cdot(\tilde n_0 {\bf u}_n)
+gn_{c0}\frac{\partial}{\partial x_{\mu}}
\delta\Gamma_{12}^{(1)}[{\bf u}_n,0],  \label{L21} \\
(\hat L_{22}{\bf u}_c)_{\mu} &=&-gn_{c0}
\frac{\partial}{\partial x_{\mu}} \bm{\nabla}\cdot (n_{c0}{\bf u}_c)
+gn_{c0}\frac{\partial}{\partial x_{\mu}}\delta\Gamma_{12}^{(1)}[0,{\bf u}_c] \label{L22}.
\end{eqnarray}
\end{subequations}
Similarly, the matrix $\bm{D}$ is block-diagonal ($\hat D_{12} = \hat
D_{21} = 0$), with elements
$\hat D_{11} = m\tilde n_0 \hat 1,
\hat D_{22} =  m n_{c0} \hat 1$.
Here, $\hat 1$ is a $3\times 3$ unit matrix. 
We note that the matrix $\bm{L}$ has the Hermitian property
\begin{equation}
\int d{\bf r}{\bf u}'\cdot (\bm{L}{\bf u})=
\int d{\bf r} {\bf u}\cdot (\bm{L}{\bf u}').
\label{hermite}
\end{equation}

The coupled equations~(\ref{eq:uc2}) and (\ref{eq:un2}) can be rewritten in terms of
the variational functional
\begin{equation}
J[{\bf u}_n,{\bf u}_c]=\frac{U[{\bf u}_n,{\bf u}_c]}{K[{\bf u}_n,{\bf u}_c]},
\end{equation}
where
\begin{equation}
U[{\bf u}_n,{\bf u}_c]\equiv\frac{1}{2}\int d{\bf r}{\bf u}\cdot(\hat L{\bf u}), ~~
K[{\bf u}_n,{\bf u}_c]\equiv\frac{1}{2}\int d{\bf r}{\bf u}\cdot(\hat D{\bf u}).
\end{equation}
Using the Hermitian property of $\bm{L}$ in Eq.~(\ref{hermite}),
one can prove that the requirement that the functional $J$ be stationary
leads to the required equations in Eqs.~(\ref{eq:uc}) and (\ref{eq:un}) and $\omega^2$ is 
identified with the stationary value of the functional $J$.
One can therefore evaluate the collective mode frequency using a variational
ansatz for ${\bf u}_n$ and ${\bf u}_c$ in the variational functional $J[{\bf u}_n,{\bf u}_c]$.


\section{General expression for damping of hydrodynamic modes}
\label{sec:damping}
In this section, we derive a general expression of hydrodynamic damping of a collective mode
due to transport coefficients.
The general expression for hydrodynamic damping of collective modes in a trapped Bose gas
was first derived in Ref.~\cite{NG} above $T_{\rm BEC}$. 

We now include hydrodynamic dissipation involving $\kappa,\eta$ and $\tau_{\mu}$
in the two-fluid equations.
Similarly to Eq.~(\ref{matrix_eq}), one can write the coupled equations for ${\bf u}_n$ and
${\bf u}_c$, which are given in Eqs.~(\ref{eq:uc}) and (\ref{eq:un}), in a matrix form
\begin{equation}
\bm{L}{\bf u}+\bm{F}{\bf u}=\omega^2\bm{D}{\bf u}.
\end{equation}
Here $\bm{F}$ represents the dissipative terms in the two-fluid equations and
has the block structure
\begin{equation}
{\bm F} = \left ( 
\begin{matrix}
 \hat F_{11}&\hat F_{12}\cr
 \hat F_{21}&\hat F_{22} \end{matrix}
 \right ),
 \label{matrixD}
\end{equation}
with the matrix elements
\begin{subequations}
\begin{eqnarray}
(\hat F_{11}{\bf u}_n)_{\mu}&=& -\frac{\partial}{\partial x_{\mu}}
\left\{\frac{2}{3}gn_{c0}(i\omega\tau_{\mu})\delta\Gamma_{12}^{(1)}[{\bf u}_n,0]
-\left(\frac{2}{3}+\frac{4n_{c0}\sigma_H\sigma_1}{9\tilde n_0}\right)
\bm{\nabla}\cdot(\kappa\bm{\nabla}\delta T_{\omega}[{\bf u}_n,0])\right\}
\label{F11} \cr
&&+\left\{i\omega\tau_{\mu}\delta\Gamma_{12}^{(1)}[{\bf u}_n,0]
-\frac{2\sigma_H\sigma_1}{3g\tilde n_0}\bm{\nabla}\cdot
(\kappa\bm{\nabla}\delta T_{\omega}[{\bf u}_n,0])\right\}\frac{\partial U_0}
{\partial x_{\mu}} \cr
&&+i\omega\frac{\partial}{\partial x_{\nu}}
2\eta\left(u_{\mu\nu}-\frac{1}{3}\bm{\nabla}\cdot{\bf u}_n\delta_{\mu\nu}\right) \\
(\hat F_{12}{\bf u}_c)_{\mu}&=&-\frac{\partial}{\partial x_{\mu}}\left\{
\frac{2}{3}gn_{c0}(i\omega\tau_{\mu})\delta\Gamma_{12}^{(1)}[0,{\bf u}_c]
-\left(\frac{2}{3}+\frac{4n_{c0}\sigma_H\sigma_1}{9\tilde n_0}\right)
\bm{\nabla}\cdot(\kappa\bm{\nabla}\delta T_{\omega}[0,{\bf u}_c])\right\} \cr
&&+\left\{i\omega\tau_{\mu}\delta\Gamma_{12}^{(1)}[0,{\bf u}_c]-
\frac{2\sigma_H\sigma_1}{3g\tilde n_0}\bm{\nabla}\cdot(\kappa\bm{\nabla}
\delta T_{\omega}[0,{\bf u}_c])\right\}\frac{\partial U_0}{\partial x_{\mu}},
 \label{F12}\\
(\hat F_{21}{\bf u}_n)&=&
gn_{c0}\frac{\partial}{\partial x_{\mu}}
\left[i\omega\tau_{\mu}\delta\Gamma_{12}^{(1)}[{\bf u}_n,0]
-\frac{2\sigma_H\sigma_1}{3g\tilde n_0}\bm{\nabla}\cdot
(\kappa\bm{\nabla} \delta T_{\omega}[{\bf u}_n,0]) \right], 
\label{F21}\\
(\hat F_{22}{\bf u}_c)&=&
gn_{c0}\frac{\partial}{\partial x_{\mu}}
\left[i\omega\tau_{\mu}\delta\Gamma_{12}^{(1)}[0,{\bf u}_c]
-\frac{2\sigma_H\sigma_1}{3g\tilde n_0}\bm{\nabla}\cdot
(\kappa\bm{\nabla} \delta T_{\omega}[0,{\bf u}_c]) \right].
 \label{F22}
\end{eqnarray}
\end{subequations}
In our subsequent analysis, we treat $\bm{F}$ as a small perturbation to the undamped equation
in Eq.~(\ref{matrix_eq}).
To find a solution for ${\bf u}_n,{\bf u}_c$
including damping, we expand the vector ${\bf u}$ in terms
of undamped normal-mode solutions ${\bf u}_{\alpha}$ ($\alpha$ is the mode
index):
\begin{equation}
{\bf u}=\sum_{\alpha}c_{\alpha}{\bf u}_{\alpha},~~
\bm{L}{\bf u}_{\alpha}=\omega_{\alpha}^2\bm{D} {\bf u}_{\alpha}.
\end{equation}
From the Hermitian property of the operator $\hat L$, one can show that
these normal-mode solutions satisfy the orthonormality relation~\cite{ZNG}
\begin{equation}
\int d{\bf r}\, {\bf u}_\alpha \cdot (\bm{D}{\bf u}_\beta) =
\delta_{\alpha \beta}\,.
\label{orthonormal}
\end{equation}
We note that in the above relation, we assume that the normal-mode solutions are normalized.
Making use of (\ref{orthonormal}), we obtain a linear equation for the coefficient
$C_{\alpha}$:
\begin{equation}
(\omega^2-\omega_{\alpha}^2)C_{\alpha}=
\int d{\bf r} {\bf u}_{\alpha}\cdot \sum_{\alpha'}C_{\alpha'}
\hat F{\bf u_{\alpha'}}\equiv\sum_{\alpha'}V_{\alpha\alpha'}C_{\alpha'},
\end{equation}
where the matrix element $V_{\alpha\alpha'}$ is defined by
\begin{equation}
V_{\alpha\alpha'}\equiv \int d{\bf r}{\bf u}_{\alpha}\cdot
\hat F{\bf u}_{\alpha'}.
\end{equation}
We note that $V_{\alpha\alpha'}$ also depend on the frequency $\omega$.

Expanding the mode frequency to first order in the perturbation $\bm{F}$
as $\omega=\omega_{\alpha}+\Delta\omega_{\alpha}$, we find
\begin{equation}
\Delta\omega_{\alpha}=\frac{1}{2\omega_{\alpha}}V_{\alpha\alpha}|_{\omega=\omega_{\alpha}},
\end{equation}
where
\begin{eqnarray}
V_{\alpha\alpha}|_{\omega=\omega_{\alpha}}&=& 
-i\omega_{\alpha} 
\int d{\bf r} \Bigl\{\frac{g\tau_{\mu}}{\sigma_H}
(\delta\Gamma_{12}^{(1)}[{\bf u}_{n\alpha},{\bf u}_{c\alpha}])^2
+\frac{\kappa}{T_0}|\bm{\nabla}\delta T_{\omega_{\alpha}}
[{\bf u}_{n\alpha},{\bf u}_{c\alpha}]|^2 \cr
&&+\frac{\eta}{2}\left(\frac{\partial u_{n\alpha\nu}}{\partial x_{\mu}}
+\frac{\partial u_{n\alpha\mu}}{\partial x_{\nu}}
-\frac{2}{3}\delta_{\mu\nu}\bm{\nabla}\cdot{\bf u}_{n\alpha}
\right)^2\Bigr\}.
\end{eqnarray}
We thus find that $\Delta\omega_{\alpha}=-i\Gamma_{\alpha}$, where the damping
rate $\Gamma_{\alpha}$ of the mode $\alpha$ is given by
\begin{eqnarray}
\Gamma_{\alpha}&=&
\int d{\bf r} \Biggl\{\frac{g\tau_{\mu}}{\sigma_H}
(\delta\Gamma_{12}^{(1)}[{\bf u}_{n\alpha},{\bf u}_{c\alpha}])^2
+\frac{\kappa}{T_0}|\bm{\nabla}\delta T_{\omega_{\alpha}}
[{\bf u}_{n\alpha},{\bf u}_{c\alpha}]|^2 \cr
&&+\frac{\eta}{2}\left(\frac{\partial u_{n\alpha\nu}}{\partial x_{\mu}}
+\frac{\partial u_{n\alpha\mu}}{\partial x_{\nu}}
-\frac{2}{3}\delta_{\mu\nu}\bm{\nabla}\cdot{\bf u}_{n\alpha}
\right)^2\Biggr\} \cr
&&\times
\left[2\int d{\bf r}m(n_{c0}u_{c\alpha}^2+\tilde n_0u_{n\alpha}^2)\right]^{-1}.
\label{gamma1}
\end{eqnarray}
Here we explicitly display the normalization factor in Eq.~(\ref{gamma1}).
This expression for the damping rate can also be written in terms of the second
viscosity coefficients:
\begin{eqnarray}
\Gamma_{\alpha}&=&
\int d{\bf r} \Biggl\{
\zeta_2(\bm{\nabla}\cdot{\bf u}_{n\alpha})^2
+2\zeta_1(\bm{\nabla}\cdot{\bf u}_{n\alpha})\bm{\nabla}\cdot
[mn_{c0}({\bf u}_{c\alpha}-{\bf u}_{n\alpha})]\cr
&&+\zeta_3\{\bm{\nabla}\cdot[mn_{c0}({\bf u}_{c\alpha}-{\bf u}_{n\alpha})]\}^2 
+\frac{\kappa}{T_0}|\bm{\nabla}\delta T_{\omega_{\alpha}}
[{\bf u}_{n\alpha},{\bf u}_{c\alpha}]|^2 \cr
&&+\frac{\eta}{2}\left(\frac{\partial u_{n\alpha\nu}}{\partial x_{\mu}}
+\frac{\partial u_{n\alpha\mu}}{\partial x_{\nu}}
-\frac{2}{3}\delta_{\mu\nu}\bm{\nabla}\cdot{\bf u}_{n\alpha}\right)^2\Biggr\} \cr
&&\times\left[2\int d{\bf r}m(n_{c0}u_{c\alpha}^2+\tilde n_0u_{n\alpha}^2)\right]^{-1}.
\label{gamma2}
\end{eqnarray}

The formula in Eq.~(\ref{gamma2}) for the damping rate can be understood
in terms of the entropy production \cite{LL}.
The local entropy production rate $R_s({\bf r},t)$ in the two-fluid equations 
\cite{Khalatnikov} is given in
Eq.~(87) of Ref.~\cite{LK}.
Assuming a normal-mode oscillation of the form
\begin{equation}
{\bf v}_n({\bf r},t)={\bf u}_{n\alpha}({\bf r})\cos\omega_{\alpha}t,~~
{\bf v}_c({\bf r},t)={\bf u}_{c\alpha}({\bf r})\cos\omega_{\alpha}t,
\end{equation}
we find that the time average of the total entropy production rate is given by
\begin{eqnarray}
\langle R_s\rangle&\equiv &\frac{\omega_{\alpha}}{2\pi}\int_0^{\frac{2\pi}{\omega_{\alpha}}}
dt\int d{\bf r}R_s({\bf r},t) \cr
&=&
\frac{1}{2}\int d{\bf r} \Biggl\{
\zeta_2(\bm{\nabla}\cdot{\bf u}_{n\alpha})^2
+2\zeta_1(\bm{\nabla}\cdot{\bf u}_{n\alpha})\bm{\nabla}\cdot
[mn_{c0}({\bf u}_{c\alpha}-{\bf u}_{n\alpha})]\cr
&&+\zeta_3\{\bm{\nabla}\cdot[mn_{c0}({\bf u}_{c\alpha}-{\bf u}_{n\alpha})]\}^2 
+\frac{\kappa}{T_0}|\bm{\nabla}\delta T_{\omega_{\alpha}}
[{\bf u}_{n\alpha},{\bf u}_{c\alpha}]|^2 \cr
&&+\frac{\eta}{2}\left(\frac{\partial u_{n\alpha\nu}}{\partial x_{\mu}}
+\frac{\partial u_{n\alpha\mu}}{\partial x_{\nu}}
-\frac{2}{3}\delta_{\mu\nu}\bm{\nabla}\cdot{\bf u}_{n\alpha}
\right)^2\Biggr\}.
\end{eqnarray}
On the other hand, the total mechanical energy is
\begin{equation}
\langle E_{\rm mech}\rangle=\frac{1}{2}\int d{\bf r}
(m\tilde n_0 u_{n\alpha}^2+mn_{c0}u_{c\alpha}^2).
\end{equation}
One can then write the damping rate $\Gamma_{\alpha}$ as
\begin{equation}
\Gamma_{\alpha}=\frac{\langle R_s \rangle}{2\langle E_{\rm mech}\rangle}.
\end{equation}
This general expression for damping was first given in the classic work 
in Landau and Lifshitz (LL)~\cite{LL} for classical fluids.
It was later used by Kavoulakis et al~\cite{KPS} to study damping in trapped Bose gases
above $T_{\rm BEC}$. 
This kind of LL damping formula was discussed in the case of superfluid $^4$He
by Wilks \cite{Wilks}.

So far we have not dealt with the problem arising from the fact that
in a trapped Bose gas, the decreasing density in the tail of the thermal
cloud always leads the breakdown of the hydrodynamic description.
As pointed out by Kavoulakis et al.~\cite{KPS}, this causes trouble
in using Eq.~(\ref{gamma1}) or (\ref{gamma2}) to evaluate the damping of modes in a
trapped Bose gas.
In Refs.~\cite{NG,KPS}, this problem was handled in a physically motivated but {\it ad hoc}
manner, by introducing a spatial cutoff in the integral.

In this paper, we propose a new, more microscopic, procedure to deal with this problem.
As we discussed in Ref.~\cite{LK}, the fact that the condensate and noncondensate atoms
are not in complete local equilibrium can be taken into account by 
introducing the frequency-dependent second viscosity coefficients
\begin{equation}
\zeta_i(\omega)=\frac{\zeta_i}{1-i\omega\tau_{\mu}}.
\end{equation}
Similarly, one can also introduce the frequency dependence in the shear 
viscosity and the thermal conductivity as
\begin{equation}
\kappa(\omega)=\frac{\kappa}{1-i\omega\tau_{\kappa}},~~
\eta(\omega)=\frac{\eta}{1-i\omega\tau_{\eta}}.
\end{equation}
In the Appendix A, we give a more detailed discussion and derivation of these
frequency-dependent transport coefficients starting from the kinetic equation.
Replacing the transport coefficients in Eq.~(\ref{gamma2}) with
$\kappa(\omega_{\alpha}),\eta(\omega_{\alpha})$ and $\zeta_i(\omega_{\alpha})$,
and taking real part, we find the damping rate of a collective mode in a trapped Bose gas
\begin{eqnarray}
\Gamma_{\alpha}&=&
\int d{\bf r} \Biggl\{\frac{1}{1+(\omega_{\alpha}\tau_{\mu})^2}
\frac{g\tau_{\mu}}{\sigma_H}
\left(\delta\Gamma_{12}^{(1)}[{\bf u}_{n\alpha},{\bf u}_{c\alpha}]\right)^2
+\frac{1}{1+(\omega_{\alpha}\tau_{\kappa})^2}
\frac{\kappa}{T_0}|\bm{\nabla}\delta T_{\omega_{\alpha}}
[{\bf u}_{n\alpha},{\bf u}_{c\alpha}]|^2 \cr
&&+\frac{1}{1+(\omega_{\alpha}\tau_{\eta})^2}
\frac{\eta}{2}\left(\frac{\partial u_{n\alpha\nu}}{\partial x_{\mu}}
+\frac{\partial u_{n\alpha\mu}}{\partial x_{\nu}}
-\frac{2}{3}\bm{\nabla}\cdot{\bf u}_{n\alpha}\delta_{\mu\nu}\right)^2\Biggr\} \cr
&&\times\left[2\int d{\bf r}m(n_{c0}u_{c\alpha}^2+\tilde n_0u_{n\alpha}^2)\right]^{-1}.
\label{gamma3}
\end{eqnarray}
We recall that $\delta T_{\omega_{\alpha}}[u_{n\alpha},u_{c\alpha}]$ and
$\delta\Gamma_{12}^{(1)}[u_{n\alpha},u_{c\alpha}]$ are defined in 
Eqs.(\ref{gamma121}) and (\ref{tomega}) in terms of the velocities $u_{n\alpha}$ and $u_{c\alpha}$.
This result in Eq.~(\ref{gamma3}) allows us to calculate the hydrodynamic damping due to various
transport processes in a trapped Bose gas (both above and below $T_{\rm BEC}$) and it is 
is the major new result of this paper.
The frequency-dependent transport coefficients in Eq.~(\ref{gamma3})
automatically yield the factors $1/[1+(\omega_{\alpha}\tau_i)]^2,~(i=\mu,\kappa,\eta)$,
which effectively introduce a spatial cutoff for $\omega_{\alpha}\tau_i>1$, i.e.,
when hydrodynamics breaks down and we enter the collisionless regime.

\section{A Uniform Bose gas}
\label{sec:uniform}
As an illustration of the physics implied by our results in Section \ref{sec:damping},
we study first and second sound in a 
uniform Bose-condensed gas using our variational expressions for the frequency
and damping.
In a dilute gas, first sound mainly involves the noncondensate oscillation
$({\bf u}_n\gg{\bf u}_c$), while second sound mainly involve the condensate
oscillation $({\bf u}_c\gg{\bf u}_n)$ (see, for example, Ref.~\cite{GZ}).
To a first approximation, we can simply use ${\bf u}_c=0$ for first sound
and ${\bf u}_n=0$ for second sound.
Using the plane-wave solution ${\bf u}_n,{\bf u}_c\propto \hat{k}\cos{\bf k}\cdot{\bf r}$
in the variational formulas, we find 
\begin{equation}
\omega_i=u_ik-i\Gamma_i~~(i=1,2),
\end{equation}
where the two sound velocities are given by
\begin{equation}
u_1^2=\frac{5\tilde P_0}{3m\tilde n_0}+\frac{2g\tilde n_0}{m}
-\frac{4gn_{c0}^2\sigma_H}{9m\tilde n_0},
\label{u1}
\end{equation}
\begin{equation}
u_2^2=\frac{gn_{c0}}{m}(1-\sigma_H),
\label{u2}
\end{equation}
and the damping rates are given by
\begin{equation}
\Gamma_1=\frac{k^2}{2m\tilde n_0}
\left[ \frac{4}{3}\eta+ \zeta_2- mn_{c0}(\zeta_1+\zeta_4)
+(mn_{c0})^2\zeta_3+\frac{4\kappa T_0}{9u_1^2}
\left(1+\frac{ 2\sigma_1\sigma_Hn_{c0} }{3\tilde n_0}\right)^2\right],
\end{equation}
\begin{equation}
\Gamma_2=\frac{k^2}{2}\left[mn_{c0}\zeta_3+
\frac{4\kappa T_0n_{c0}}{9u_2^2m\tilde n_0^2}(\sigma_1\sigma_H)^2\right].
\end{equation}
These results agree with Eqs.~(C11)-(C14) of Appendix C in Ref.~\cite{LK},
which were directly derived from the Landau-Khalatnikov two-fluid equations for a dilute Bose gas. 
We note that the two sound velocities $u_1$ and $u_2$ are slightly different from
those given in Ref.~\cite{GZ}, because Ref.~\cite{GZ} worked in the limit $\omega\tau_{\mu}\gg 1$,
and neglected the source term $\delta\Gamma_{12}$.
As shown in Ref.~\cite{ZNG}, however, these differences are quantitatively very small in the
case of a weakly-interacting Bose gas.

\section{Monopole-quadrupole mode in a degenerate normal Bose gas}
\label{sec:aboveTc}
Let us now consider collective modes in a trapped noncondensed Bose gas above
$T_{\rm BEC}$.
Here we consider the $m=0$ monopole-quadrupole collective mode in an
an axi-symmetric trap ($\omega_x=\omega_y=\omega_{\perp}\neq \omega_z$).
This type of collective mode above $T_{\rm BEC}$
was first observed in the pioneering MIT experiment \cite{MIT}, but the density
was not large enough to probe the hydrodynamic regime (see, however, discussions in
Ref.~\cite{relax}).
More recently, however, ENS experiments with metastable He$^*$ atoms studied the $m=0$ mode in
a high-density thermal cloud~\cite{ENS3}. 

Above $T_{\rm BEC}$, the Hartree-Fock mean field is negligible and thus the equilibrium
density is simply given by Eq.~(\ref{ntilde0}) with $z_0({\bf r})=e^{\beta[\mu_0-U_{\rm ext}({\bf r})]}$.
The chemical potential $\mu_0$ is determined as a function of the temperature through
$N=(k_{\rm B}T/\hbar\bar\omega)^3g_3(z_0)$, where 
$\bar\omega\equiv(\omega_x\omega_y\omega_z)^{1/3}$.
The hydrodynamic modes in a trapped Bose gas with the $m=0$ symmetry
were first discussed by Griffin, Wu, and Stringari \cite{GWS}.
The two normal-mode frequencies are temperature-independent, and are given by \cite{GWS}
\begin{equation}
\Omega_{\pm}^2=\frac{1}{3}[5\omega_{\perp}^2+4\omega_z^2
\pm\sqrt{25\omega_{\perp}^4+16\omega_z^4-32\omega_z^2\omega_{\perp}^2}].
\label{freq_hydro}
\end{equation}
The corresponding velocity field is given by
\begin{equation}
{\bf u}_n=(ax,ay,bz),
\label{eq_un}
\end{equation}
where the coefficients $a$ and $b$ satisfy the following relations:
\begin{equation}
\frac{a_{\pm}}{b_{\pm}}=\left(\frac{3\Omega^2_{\pm}}{4\omega_z^2}-2\right),~~
{\rm or}~~
\frac{b_{\pm}}{a_{\pm}}=\left(\frac{3\Omega^2_{\pm}}{2\omega_{\perp}^2}-5\right).
\label{eq_vmode}
\end{equation}

Kavoulakis et al. ~\cite{KPS} discussed the hydrodynamic damping of this $m=0$ mode
using the LL formula, with a spatial cutoff to deal with the crossover from the hydrodynamic to
collisionless regime in the tail of the thermal cloud.
In contrast, we calculate the damping rate using Eq.~(\ref{gamma3}), which eliminate the collisionless
regime in the tail of the thermal cloud through the use of the frequency-dependent transport
coefficients.
Above $T_{\rm BEC}$, there is no contribution from the second viscosity transport coefficients.
Moreover, in the $m=0$ mode above $T_{\rm BEC}$,
one can show $\bm{\nabla}T=0$ \cite{GWS,KPS} and thus the thermal conductivity
makes no contribution to the damping of the $m=0$ mode.
Thus, only the shear viscosity in Eq.~(\ref{gamma3}) contributes to the damping 
of this low-frequency collective mode.
Using Eq.~(\ref{eq_un}), we find (see also Ref.~\cite{KPS})
\begin{equation}
\Gamma_-=\frac{\displaystyle \frac{2}{3}(a_--b_-)^2\int d{\bf r}
\frac{\eta({\bf r})}{1+[\Omega_-\tau_{\eta}({\bf r})]^2}}
{\displaystyle \int d{\bf r}~ 
[a_-^2(x^2+y^2)+b_-^2 z^2]m\tilde n_0({\bf r})}.
\label{gammam1}
\end{equation}

Making use of the fact that the equilibrium density profile given by Eq.~(\ref{ntilde0}) is 
a function of $U_{\rm ext}({\bf r})$,
one can rewrite Eq.~(\ref{gammam1}) as
\begin{equation}
\Gamma_-=\frac{\displaystyle \frac{2}{3}(a_--b_-)^2
\int \frac{\eta({\bf r})}{1+[\Omega_-\tau_{\eta}({\bf r})]^2}}
{\displaystyle \left(\frac{2a_-^2}{\omega_{\perp}^2}+\frac{b_-^2}{\omega_z^2}\right)
\frac{m}{3} \int [\omega_{\perp}^2(x^2+y^2)+\omega_z^2z^2]\tilde n_0({\bf r})}.
\label{gammam2}
\end{equation}
In Eq.~(\ref{gammam2}), the factor involving the coefficients $a_-$ and $b_-$ can
be written in a simple form in terms of the undamped frequencies $\Omega_-$
and $\Omega_+$ as
\begin{eqnarray}
\frac{\displaystyle \frac{2}{3}(a_-^2-b_-^2)}
{\displaystyle \left(\frac{2a_-^2}{\omega_{\perp}^2}+\frac{b_-^2}{\omega_z^2}\right)}
&=&\frac{\displaystyle \frac{2}{3}\left(\frac{a_-}{b_-}-1\right)
\left(1-\frac{b_-}{a_-}\right)}
{\displaystyle \left(\frac{2a_-}{\omega_{\perp}^2b_-}+\frac{b_-}{\omega_z^2a_-}\right)}
=\frac{(\Omega_-^2-4\omega_z^2)(\Omega_-^2-4\omega_{\perp}^2)}
{4(5\omega_{\perp}^2+4\omega_z^2-3\Omega_-^2)} \cr
&=&\frac{(\Omega_-^2-4\omega_z^2)(\Omega_-^2-4\omega_{\perp}^2)}
{2(\Omega_{+}^2-\Omega_-^2)}.
\end{eqnarray}
Here we have used Eq.~(\ref{eq_vmode}) and Eq.~(\ref{freq_hydro}). 
We thus obtain the following simple expression for the damping rate:
\begin{equation}
\Gamma=\frac{\tilde \tau}{2(\Omega^2_+-\Omega^2_-)}
(\Omega_-^2-4\omega_z^2)(\Omega_-^2-4\omega^2_{\perp}),
\label{gamma_hydro}
\end{equation}
where we have introduced a new relaxation time [see Eq.~(\ref{eq:eta})]:
\begin{equation}
\tilde\tau\equiv
\frac{\displaystyle \int d{\bf r}\frac{\tilde P_0({\bf r})\tau_{\eta}({\bf r})}
{1+(\Omega_-\tau_{\eta})^2}}
{\displaystyle \frac{m}{3}\int d{\bf r}
[\omega_{\perp}^2(x^2+y^2)+\omega_z^2z^2]\tilde n_0({\bf r})}.
\label{damp_hydro}
\end{equation}
Here we have used the relation [see Eqs.~(\ref{ntilde0}), (\ref{ptilde0}) and (\ref{eq:eta})]
$\eta({\bf r})=\tilde P_0({\bf r}) \tau_{\eta}({\bf r})$.
Using the equilibrium relation ${\bf \nabla}\tilde P_0+\tilde n_0{\bf \nabla}U_{\rm ext}=0$,
which is valid when neglecting the HF mean field, one can reduce
Eq.~(\ref{damp_hydro}) to
\begin{equation}
\tilde\tau=\frac{\displaystyle \int d{\bf r}\frac{\tilde P_0({\bf r})\tau_\eta}
{1+(\Omega_-\tau_{\eta})^2}}{\displaystyle \int d{\bf r}\tilde P_0({\bf r})}.
\label{damp_hydro2}
\end{equation}

Before presenting results given by an explicit evaluation of the relaxation time $\tilde\tau$ in
Eq.~(\ref{damp_hydro}) using the parameters for the ENS trap~\cite{ENS3},
it is useful to comment on the relation between the present calculation of
hydrodynamic damping and the moment method developed by Gu\'ery-Odelin et al. \cite{moment}.
These authors applied the moment method to the classical-gas Boltzmann equation.
It is straightforward to generalize their method to the
kinetic equation for a Bose-degenerate gas (see, for example, Ref.~\cite{Nikuni}).
This moment method is briefly reviewed in Appendix B, and here we simply give
to the final results.
In the moment method, collisions are characterized by a single parameter, the
quadrupole relaxation time $\tau$ defined in Eq.~(\ref{tau_q}).
In the hydrodynamic limit $\omega_z\tau\ll 1$, the moment method reproduces the hydrodynamic
frequency in Eq.~(\ref{freq_hydro}), but the damping is now given by Eq.~(\ref{gamma_moment}).
The moment result for the damping has the same form as
Eq.~(\ref{gamma_hydro}), except that $\tau$ replaces $\tilde\tau$.
Both $\tilde\tau$ and $\tau$ are related to the same position-dependent viscous relaxation time
$\tau_{\eta}({\bf r})$, but involve different spatial averages
[see Eq.~(\ref{damp_hydro}) and Eq.~(\ref{tau_q})].

\begin{figure}
\centerline{\epsfig{file=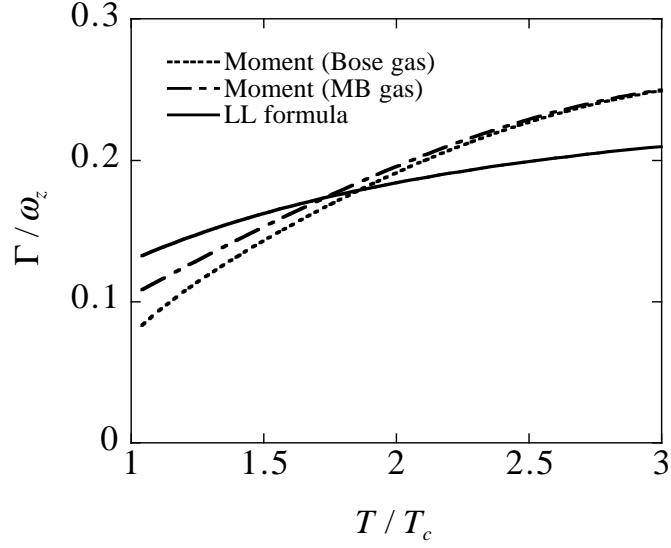,height=3.0in}}
\caption{Temperature dependence of the hydrodynamic damping rate of the $m=0$
mode in a degenerate Bose gas above $T_{\rm BEC}$, calculated from
Eq.~(\ref{gamma_hydro}).
We also show the result obtained from the moment method by the broken line 
[see Fig.~\ref{moment3}(a)].
For comparison, we also plot the moment result for the Maxwell-Boltzmann (MB) gas.
}
\label{damping}
\end{figure}

Our evaluation of the damping $\Gamma$ in Eq.~(\ref{gamma_hydro}) is based on the ENS trap parameters
\cite{ENS2,ENS3}:
trap frequencies $\omega_{\perp}/2\pi=988,\omega_z/2\pi=115$,
total number of atoms $N=8.2\times 10^6$, and $s$-wave scattering
length $a=16$nm.
The ideal Bose gas transition temperature is given by
$T_c=(\hbar\bar\omega/k_{\rm B})(N/1.202)^{1/3}=4.39\mu$K.
In the temperature region
$T_c < T < 3 T_c$, our calculations show that $\omega_z\tau_{\eta}({\bf r}=0)\ll 1$.
Thus the dominant contribution in the integral of Eq.~(\ref{damp_hydro}) 
arises from the low density tail of the cloud where $\omega_z\tau_\eta({\bf r}) \sim 1$.
In Fig.~\ref{damping}, we plot the temperature dependence of the damping rate.
For comparison, we also plot the damping calculated using the moment method.
We find that the two methods give results of the same order of magnitude, but there are
significant differences.

\begin{figure}
\centerline{\epsfig{file=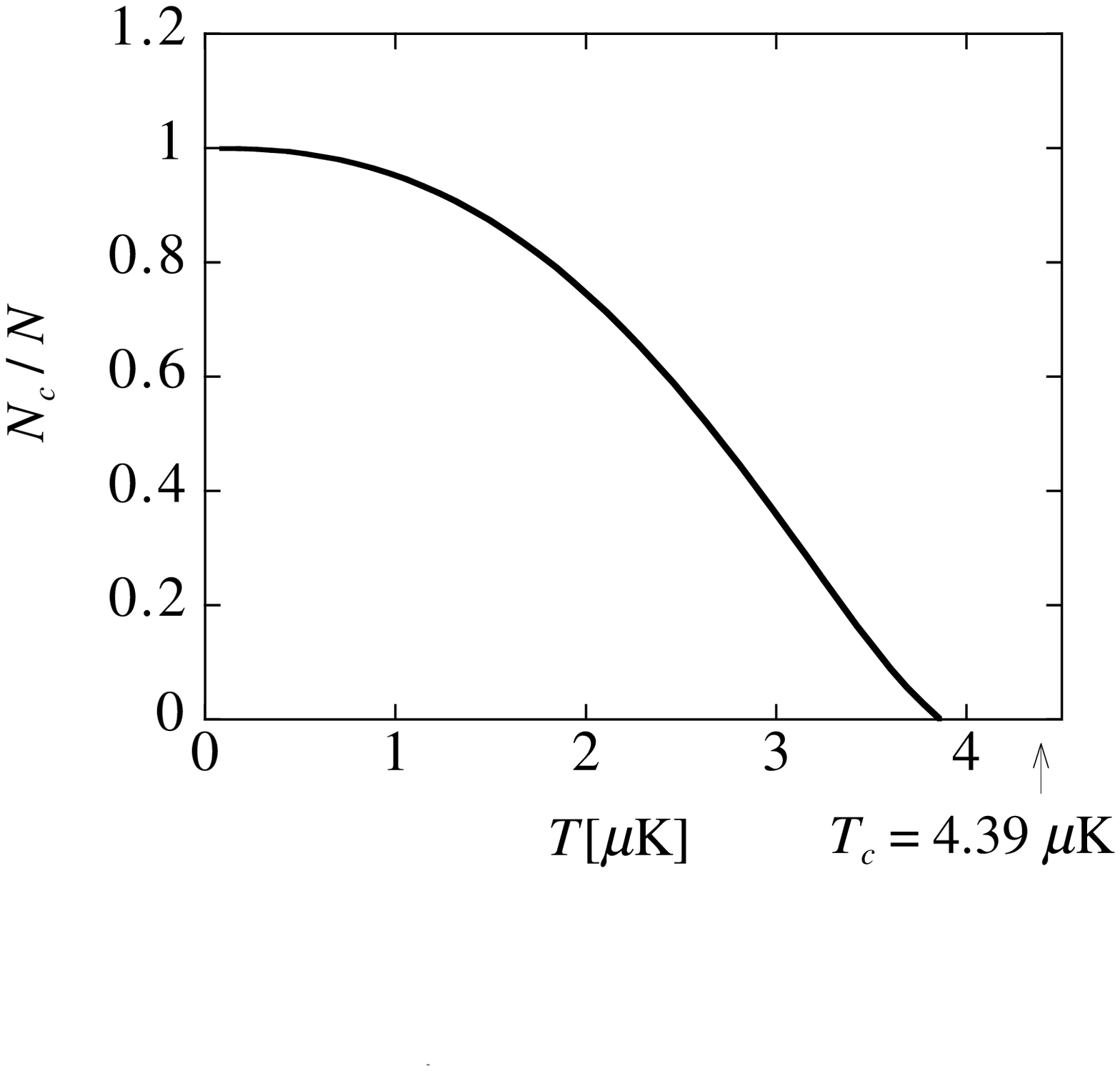,height=3.5in}}
\caption{Condensate fraction versus temperature.}
\label{fraction}
\end{figure}

\begin{figure}
\centerline{\epsfig{file=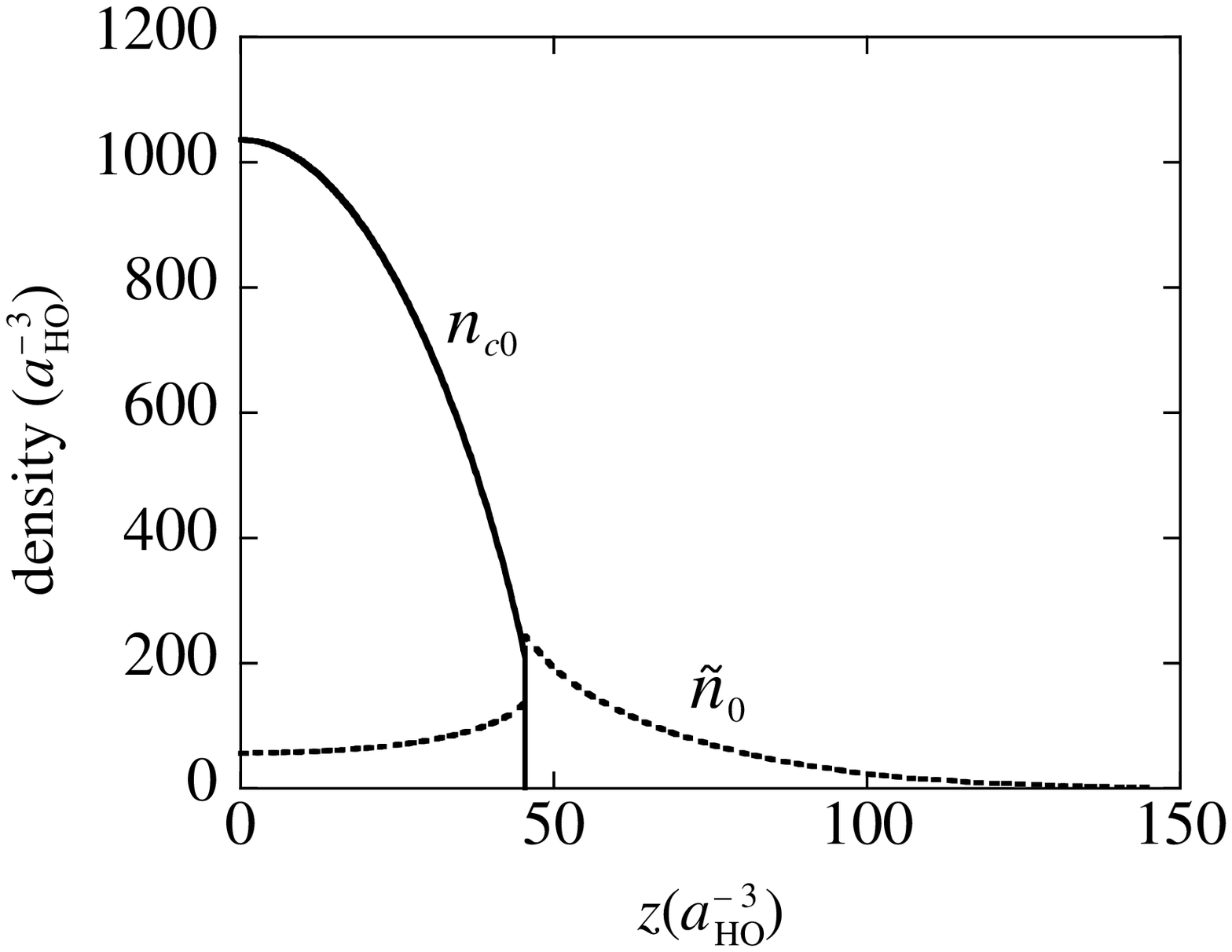,height=3.0in}}
\caption{Density profile of the condensate and noncondensate
along the $z$ axis at $T=3\mu$K ($\simeq 0.68 T_{c}$).
The length unit is the average harmonic oscillator length
$a_{\rm HO}\equiv \hbar/m\bar\omega$, and the density unit
is $a_{\rm HO}^{-3}$.
The discontinuous change in $n_{c0}$ and $\tilde n_0$ at the condensate boundary
is a well-known artifact arising from using the Thomas-Fermi approximation.}
\label{density}
\end{figure}

\section{Monopole-quadrupole mode in a Bose-condensed gas}
\label{sec:belowTc}
In this section, we consider the $m=0$ collective mode in the superfluid phase below
$T_{\rm BEC}$.
For this purpose, we first need to calculate various equilibrium quantities, solving Eq.~(\ref{nc0}) 
and Eq.~(\ref{ntilde0}) self-consistently.
Fig. \ref{fraction} shows the temperature dependence of the condensate fraction.
The effective Bose condensation temperature is lower than the free gas result $T_c$,
due to the mean field.
Fig. \ref{density} shows the associated density profile of the condensate and noncondensate
components at $T=3\mu K\simeq 0.68 T_c$.

Below $T_{\rm BEC}$, the gas in general exhibits coupled oscillations of the condensate
and noncondensate components.
At finite temperatures slightly below $T_{\rm BEC}$, one has a ``condensate mode",
in which the condensate component mainly oscillates, and a
``noncondensate mode", in which the noncondensate component mainly oscillates \cite{ZNG}.
As one might expect, the frequencies of the modes are close to those of a pure condensate mode
at $T=0$ and a pure noncondensate mode above $T_{\rm BEC}$, respectively.
These frequencies are slightly shifted due to coupling between the two components.
In the calculations in this Section, we focus entirely on the damping of the modes, and neglect 
these relatively small frequency shifts.
Thus, we neglect the condensate oscillation in the
noncondensate mode and noncondensate oscillation in the
condensate mode.
We calculate the hydrodynamic damping of these two modes.
In contrast to the moment method results discussed in Appendix B, our present results
are only valid in the hydrodynamic regime.

\subsection{Noncondensate mode}
\begin{figure}
\centerline{\epsfig{file=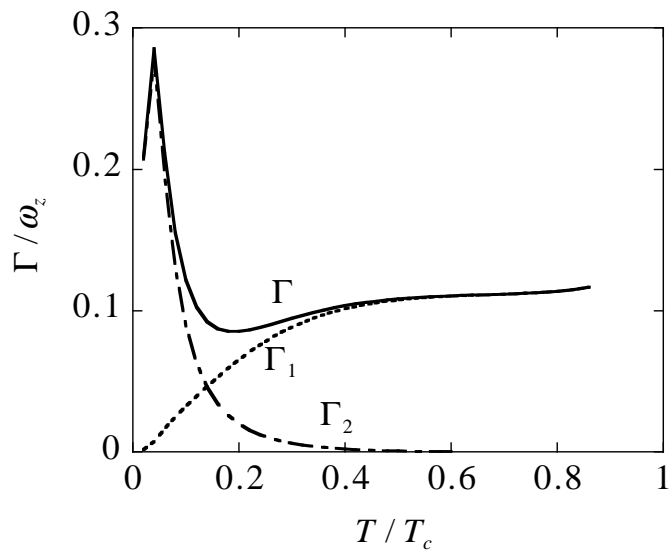,height=3.0in}}
\caption{Damping rate of the noncondensate mode below the BEC transition
temperature.
The broken lines gives the separate contributions from the shear viscosity
($\Gamma_1$) and the second viscosity ($\Gamma_2$).
Compare with results above $T_c$ shown in Fig.~1.}
\label{NCdamping}
\end{figure}

We first consider damping of the noncondensate mode of monopole-quadrupole symmetry.
In Eq.~(\ref{gamma1}), we use ${\bf u}_n$ given by Eq.~(\ref{eq:un}) 
Eq.~(\ref{eq_un}) with Eq.~(\ref{eq_vmode}) and take ${\bf u}_c=0$.
For simplicity, we assume that $\bm{\nabla}T=0$ also holds below $T_{\rm BEC}$.
We then find that the damping consists of two contributions, $\Gamma=\Gamma_1+\Gamma_2$,
where $\Gamma_1$ is the contribution from the shear viscosity, again given by 
Eq.~(\ref{gamma_hydro}), while $\Gamma_2$ is due to the second viscosity, which is given by
\begin{equation}
\Gamma_2=\frac{\displaystyle \int d{\bf r} \frac{1}{1+(\Omega_-\tau_{\mu})^2}
\frac{g\tau_{\mu}}{\sigma_H}\left(\delta\Gamma_{12}^{(1)}[{\bf u}_n,0]\right)^2}
{2m \int d{\bf r}\tilde n_0 u_n^2}
\end{equation}
As noted above, the frequency $\Omega_-$ of this mode is well approximated by $\Omega_-$
as given in Eq.~(\ref{freq_hydro}) for $T>T_c$.
In Fig.~\ref{NCdamping}, we plot the temperature dependence of the damping of the
noncondensate mode below $T_{\rm BEC}$.
The contribution from the shear viscosity is dominant for
$T\gtrsim 0.3 T_c$, with the contribution from the second viscosity coefficient only taking
over at very low temperatures.

\subsection{Condensate mode}
We next consider the condensate mode below the superfluid transition temperature.
The pure condensate mode frequencies at $T=0$ were first given by Stringari \cite{Stringari}:
\begin{equation}
\Omega_{\pm}^2=2\omega_{\perp}^2+\frac{3}{2}\omega_z^2\pm
\frac{1}{2}\sqrt{16\omega_{\perp}^4+9\omega_z^4-16\omega_{\perp}^2\omega_z^2}.
\label{freq_c}
\end{equation}
The associated condensate velocity field is given by
\begin{equation}
{\bf u}_c=(ax,ay,bz),
\label{velocity_c}
\end{equation}
where one finds
\begin{equation}
\frac{a_{\pm}}{b_{\pm}}=\left(\frac{\Omega^2_{\pm}}{2\omega_z^2}-\frac{3}{2}\right),~~{\rm or}~~
\frac{b_{\pm}}{a_{\pm}}=\left(\frac{\Omega^2_{\pm}}{\omega_{\perp}^2}-4\right)a_{\pm}.
\end{equation}
Using this pure condensate mode solution for ${\bf u}_c$,
and setting the noncondensate velocity ${\bf u}_n=0$ and ignoring any temperature fluctuation 
$\delta T$,
we obtain a simple formula from Eqs.~(\ref{gamma3}) and (\ref{gamma121}) for
the damping of the low-frequency condensate mode,
\begin{equation}
\Gamma_-=\frac{\displaystyle \int d{\bf r}\frac{\displaystyle 
g\tau_{\mu}\sigma_H}{1+(\Omega_-\tau_{\mu})^2}
[\bm{\nabla}\cdot(n_{c0}{\bf u}_c)]^2}{\displaystyle 2m\int d{\bf r}n_{c0}u_c^2}.
\label{gamma_c}
\end{equation}
Here $\Omega_-$ is the frequency given by Eq.~(\ref{freq_c}) and is approximately
$\Omega_-\approx \sqrt{\frac{5}{2}}\omega_z$, while ${\bf u}_c$ is given by
Eq.~(\ref{velocity_c}), both of these describing undamped hydrodynamic mode.

The expression for the damping rate in Eq.~(\ref{gamma_c}) which has been found here
in the hydrodynamic regime is similar to Eq.~(79)
of Ref.~\cite{Nikuni}, which gives the collisional damping of condensate collective modes
in the collisionless regime.
In fact, one can show that in the limit $\Omega_-\tau_{\mu}\gg 1$, Eq.~(\ref{gamma_c})
formally reduces precisely to the expression in Eq.~(79) of Ref.~\cite{Nikuni}.
As shown in Ref.~\cite{Nikuni}, this latter expression for the condensate mode damping
is equivalent to the result derived by the method of Williams and Griffin \cite{WG}
in the collisionless regime, under the assumption that the thermal cloud always remained in
static thermal equilibrium.
This makes sense since our assumptions of ${\bf u}_n=0$ and $\bm{\nabla}T=0$ are equivalent
to assuming a static thermal cloud.

\begin{figure}
\centerline{\epsfig{file=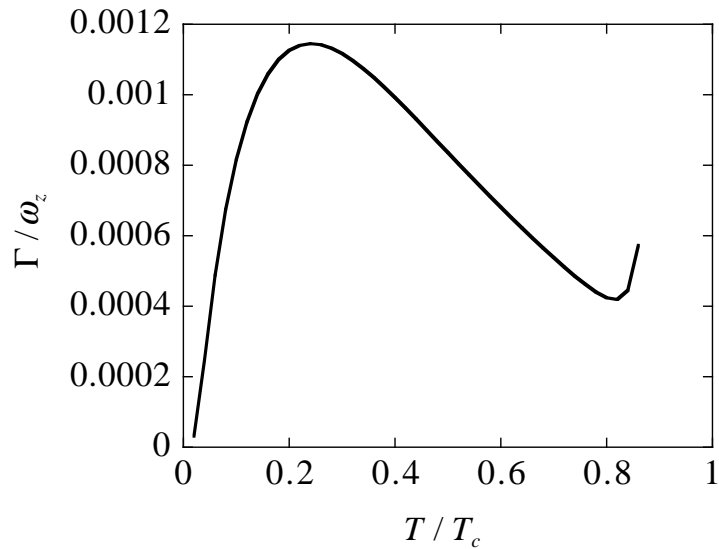,height=3.0in}}
\caption{Damping rate of the $m=0$ condensate mode.}
\label{Cdamping}
\end{figure}

In Fig.~\ref{Cdamping}, we plot the temperature dependence of the damping of this
condensate mode.
The damping of this condensate mode is extremely small, simply because we are always
in the extreme hydrodynamic limit $\omega_z\tau_{\mu}\ll 1$. 

\section{Discussion and Concluding Remarks}
Recently, the Landau-Khalatnikov two-fluid equations were derived \cite{CJP,LK} for a trapped
Bose gas in the collision-dominated local equilibrium domain.
These involve the transport coefficients \cite{relax} (thermal conductivity, shear viscosity,
and four second viscosity coefficients) describing the processes leading to local equilibrium
in a superfluid, from which various relaxation times $\tau_i$ can be extracted.
In the present paper, we used these equations to derive a general expression for the
damping of hydrodynamic modes in a trapped Bose gas.
This formula makes use of the variational solution of the two-fluid hydrodynamic equations
in the Landau limit ($\omega\tau_i\ll1$, where $\omega$ is the frequency of a hydrodynamic mode).
We hope that our work will stimulate further experimental studies of the collision-dominated
hydrodynamic regime in trapped Bose gases.

As illustration, we used our formalism to evaluate the hydrodynamic damping of the $m=0$
monopole-quadrupole mode of a cigar-shaped trap, using parameters appropriate to the recent
ENS experiments \cite{ENS1,ENS2,ENS3} on metastable $^4$He$^*$.
We presented results for both $T>T_{\rm BEC}$ and $T<T_{\rm BEC}$.
We also used the moment method developed by Gu\'ery-Odelin et al. \cite{moment} for a trapped
classical gas and give results (see Appendix B) for a degenerate trapped Bose gas above $T_{\rm BEC}$.
The advantage of the moment approach \cite{moment} is that the resulting equations of motion for
various moments can be solved over the entire frequency domain,
including both the collisionless ($\omega\tau>1$) and hydrodynamic regions ($\omega\tau<1$).

For temperatures characteristic of the ENS experiments ($T\sim 3T_c$), the effect
of Bose statistics is almost negligible.
Thus, as expected, our calculated values of the damping of the coupled $m=0$ 
monopole-quadrupole mode (due entirely to the shear viscosity) are in good agreement with
the moment calculations of Ref.~\cite{moment} for a classical gas.
Indeed, our calculation shows that the difference remains small down to the superfluid transition $T_c$.
However, as discussed in Ref.~\cite{ENS2,ENS3}, the analysis of the ENS experimental data exhibits
a puzzling discrepancy.
The damping and frequency of the $m=0$ mode is consistent with a maximum value of the collision
rate [defined in Eq.~(\ref{taucl})] being given by $\Gamma_{\rm coll}\simeq 2\times 10^{-1}$s$^{-1}$,
which is achieved at $T=3T_c$.
Lowering the temperature led to an apparent {\it decrease} in the values of $\Gamma_{\rm coll}$,
the latter being determined by the measured changes in the frequency and damping of the $m=0$ mode.
This result seems inconsistent with the calculated value of $\Gamma_{\rm coll}=\sqrt{2}\bar n
\sigma \bar v$ at $T_c$, using the measured values at $T_c$ of the average
density and velocity of atoms, and the collision cross-section $\sigma=8\pi a^2$.
Our calculated value is 
$\Gamma_{\rm coll}(T_c)=9\times 10^3$s$^{-1}$ (see Fig.~\ref{moment2}), 
considerably larger than the value $\Gamma_{\rm coll}=10^3$s$^{-1}$ as
estimated from the frequency and damping of the $m=0$ mode \cite{ENS2,ENS3}.

Thus, the ENS experiment on the $m=0$ collective mode do not seem to be able to enter
deeply into the hydrodynamic regime.
Ref.~\cite{ENS2} tentatively interprets this to being due to increasing inelastic collision
processes, which effectively lead to a decreasing collision rate $\Gamma_{\rm coll}$ for
$T$ below $3T_c$.
However this does not explain why the estimated value of $\Gamma_{\rm coll}$ calculated at $T_c$
is an order of magnitude larger, which would correspond to the $m=0$ mode being deeply
into the collision-dominated hydrodynamic domain.

In order to clarify this puzzling behavior above $T_c$, it would be useful to measure the damping
and frequency of the monopole-quadrupole collective mode in the Bose-condensed region.
As discussed in Ref.~\cite{relax}, the effective value of the relaxation time $\tau_{\eta}$
associated with the shear viscosity is calculated to become much smaller as one goes below $T_c$.
This is simply because $1/\tau_{\eta}$ of the thermal cloud atoms is dominated by collisions
with condensate atoms.
As a result, even if one has a low thermal cloud density $\tilde n_0$ (spatially averaged)
such that one is in the collisionless region above $T_c$, one is automatically deep inside
the hydrodynamic region for $T<T_c$ because of the rapid build up of the condensate density
in the center of the trap.
This suggested variant of the ENS experiments effectively uses the formation of the high-density
Bose condensate to increase the collision rate which determines the frequency and damping
of the monopole-quadrupole mode.

In Section VII, we presented the first explicit calculations of the hydrodynamic damping of the
monopole-quadrupole mode in the superfluid phase.
As can be seen in Fig.~\ref{NCdamping}, the damping of this mode involving the noncondensate 
thermal cloud is still dominated by the shear viscosity down to $T\sim 0.3 T_c$.
Comparing the results above (Fig.~\ref{damping}) and below $T_c$ (Fig.~\ref{NCdamping}),
one sees the mode damping $\Gamma$ is fairly smooth going through the transition,
with only a slight decrease in magnitude below $T_c$.
This slight decrease in the value of $\Gamma$ hides the fact that (as discussed above) the 
effective value of the shear collision relaxation time $\tau$ is rapidly decreasing
as we go below $T_c$, putting one deep into the hydrodynamic domain.

In Section VII, we also evaluated the hydrodynamic damping of the monopole-quadrupole mode
in the condensate.
As seen in Fig.~\ref{Cdamping}, these damping is extremely small since one is effectively in the
``Landau limit'', $\omega\tau_{\mu}\ll 1$. 

\begin{center}
{\bf ACKNOWLEDGMENTS}
\end{center}
We thank Eugene Zaremba for his important contributions to the
initial phase of this research and also for critical comments on the final manuscript.
We also acknowledge useful discussions with Michelle Leduc.
A.G. thanks NSERC for research support.


\begin{appendix}
\section{frequency-dependent transport coefficients}
In this Appendix, we give some details on the frequency-dependent coefficients
of the shear viscosity and thermal conductivity, starting from the kinetic equation
for the noncondensate atoms.
As in the usual Chapman-Enskog procedure described in Ref.~\cite{LK}, we insert
the local equilibrium distribution function $f_{\rm leq}$ in the left hand side
of the kinetic equation, where $f_{\rm leq}$ is given by
\begin{equation}
f_{\rm leq}({\bf r},{\bf p},t)=\frac{1}{z^{-1}({\bf r},t)
e^{\beta({\bf r},t)[{\bf p}-m{\bf v}_n({\bf r},t)]^2/2m}-1},
\label{f_leq}
\end{equation}
where $z({\bf r},t)=e^{\beta({\bf r},t)[\tilde\mu({\bf r},t)-U({\bf r},t)]}$ is 
the local fugacity.
As shown in Appendix A of Ref.~\cite{LK}, the kinetic equation is then given by
\begin{eqnarray}
&&\frac{\partial f}{\partial t}
+\Biggl[\frac{1}{z}\frac{{\bf p}}{m}\cdot\bm{\nabla}z
+\frac{({\bf p}-m{\bf v}_n)^2}{2mk_{\rm B}T^2}\frac{{\bf p}}{m}\cdot\bm{\nabla}T
+\frac{{\bf p}-m{\bf v}_n}{k_{\rm B}T}\cdot\left(\frac{{\bf p}}{m}\cdot\bm{\nabla}\right)
{\bf v}_n \cr
&&+\frac{\bm{\nabla}U}{mk_{\rm B}T}\cdot({\bf p}-m{\bf v}_n)\Biggr]f_{\rm leq}(1+f_{\rm leq})
=C_{12}+C_{22}.
\label{kineq_leq}
\end{eqnarray}
In contrast to Ref.~\cite{LK}, we keep the time derivative of $f$ explicitly.
Since we are interested in small-amplitude collective oscillations, in the following
we always expand the theory to first order in the fluctuations around static equilibrium .

We first consider the shear viscosity, which is associated with the anisotropic
pressure tensor. In a linearized theory, this is given by
\begin{equation}
P_{\mu\nu}'\equiv P_{\mu\nu}-\delta_{\mu\nu}\tilde P
=\int\frac{d{\bf p}}{(2\pi\hbar)^3}
\frac{1}{m}
\left(p_{\mu}p_{\nu}-\frac{1}{3}\delta_{\mu\nu}p^2\right)
f({\bf r},{\bf p},t).
\label{linear_pmunu}
\end{equation}
The equation of motion for $P_{\mu\nu}'$ can be obtained by taking moment
of Eq.~(\ref{kineq_leq}) and linearizing it around static thermal equilibrium.
One finds that the term contributes to this moment is
\begin{equation}
\int \frac{d{\bf p}}{(2\pi\hbar)^3}\frac{1}{m}
\left(p_{\mu}p_{\nu}-\frac{1}{3}\delta_{\mu\nu}p^2\right)
\frac{{\bf p}}{k_{\rm B}T_0}\cdot \left(\frac{{\bf p}} {m} \cdot \bm{\nabla}\right)
{\bf v}_n f_0(1+f_0)=
\tilde P_0\left(\frac{\partial v_{n\nu}}{\partial x_{\mu}}+
\frac{\partial v_{n\mu}}{\partial x_{\nu}}
-\frac{2}{3}\delta_{\mu\nu}\bm{\nabla}\cdot{\bf v}_n \right).
\end{equation}
One thus obtains
\begin{equation}
\frac{\partial P'_{\mu\nu}}{\partial t}+
\tilde P_0\left(\frac{\partial v_{n\nu}}{\partial x_{\mu}}+
\frac{\partial v_{n\mu}}{\partial x_{\nu}}
-\frac{2}{3}\delta_{\mu\nu}\bm{\nabla}\cdot{\bf v}_n \right)
=\left\langle
\frac{1}{m}\left(p_{\mu}p_{\nu}-\frac{1}{3}p^2\delta_{\mu\nu}\right)
\right\rangle_{\rm coll}.
\label{eq:pmunu}
\end{equation}
where
\begin{equation}
\left\langle
\frac{1}{m}\left(p_{\mu}p_{\nu}-\frac{1}{3}\delta_{\mu\nu}p^2\right)
\right\rangle_{\rm coll}\equiv 
\int\frac{d{\bf p}}{(2\pi\hbar)^3}
\frac{1}{m}\left(p_{\mu}p_{\nu}-\frac{1}{3}\delta_{\mu\nu}p^2\right)
(C_{12}+C_{22}).
\label{pmunu_coll}
\end{equation}

The collisional contribution on the right hand side of Eq.~(\ref{eq:pmunu}) arises
from deviation of the distribution $f$ from the local equilibrium solution in Eq.~(\ref{f_leq}).
Following the Chapman-Enskog procedure, we use the ansatz $f=f_{\rm leq}+\delta f$,
where
\begin{equation}
\delta f = \sum_{\mu\nu}B_{\mu\nu}\left(p_{\mu}p_{\nu}-\frac{1}{3}\delta_{\mu\nu}p^2
\right)
f_0(1+f_0),
\label{deltafB}
\end{equation}
with $B_{\mu\nu}$ being some momentum-independent symmetric tensor.
The relation between $B_{\mu\nu}$ and $P'_{\mu\nu}$ can be found by using Eq.~(\ref{deltafB})
in Eq.~(\ref{linear_pmunu}) and carrying out the momentum integral:
\begin{eqnarray}
P_{\mu\nu}'&=&\sum_{\mu'\nu'}B_{\mu'\nu'}
\int\frac{d{\bf p}}{(2\pi\hbar)^3}\frac{1}{m}
\left(p_{\mu}p_{\nu}-\frac{1}{3}\delta_{\mu\nu}p^2\right)
\left(p_{\mu'}p_{\nu'}-\frac{1}{3}\delta_{\mu'\nu'}p^2\right)
f_0(1+f_0) \cr
&=&\frac{1}{5}\left(B_{\mu\nu}-\frac{1}{3}\delta_{\mu\nu}{\rm Tr}B\right)
\sum_{\mu'\nu'}
\int\frac{d{\bf p}}{(2\pi\hbar)^3}\frac{1}{m}
\left(p_{\mu'}p_{\nu'}-\frac{1}{3}\delta_{\mu'\nu'}p^2\right)^2 \cr
&=&
2mk_{\rm B}T_0\tilde P_0
\left(B_{\mu\nu}-
\frac{1}{3}\delta_{\mu\nu}{\rm Tr}B\right).
\label{PandB}
\end{eqnarray}
Using Eq.~(\ref{deltafB}) in Eq.~(\ref{pmunu_coll}), we find
\begin{eqnarray}
&&\left\langle
\frac{1}{m}\left(p_{\mu}p_{\nu}-\frac{1}{3}\delta_{\mu\nu}p^2
\right)
\right\rangle_{\rm coll}=
\sum_{\mu'\nu'}B_{\mu'\nu'}\int\frac{d{\bf p}}{(2\pi\hbar)^3}
\frac{1}{m}\left(p_{\mu}p_{\nu}-
\frac{1}{3}\delta_{\mu\nu}p^2\right)
\hat L\left[p_{\mu'}p_{\nu'}-\frac{1}{3}\delta_{\mu'\nu'}p^2\right] \cr
&&=\frac{1}{5}\left(B_{\mu\nu}-\frac{1}{3}\delta_{\mu\nu}{\rm Tr}B\right)
\sum_{\mu'\nu'}
\int\frac{d{\bf p}}{(2\pi\hbar)^3}\frac{1}{m}
\left(p_{\mu'}p_{\nu'}-\frac{1}{3}\delta_{\mu'\nu'}p^2\right)
\hat L\left[p_{\mu'}p_{\nu'}-\frac{1}{3}\delta_{\mu'\nu'}p^2\right],
\label{pcoll}
\end{eqnarray}
where $\hat L$ is the linearized collision operator defined in
Eqs.~(46) and (48) of Ref.~\cite{LK}.
Combining Eq.~(\ref{PandB}) and Eq.~(\ref{pcoll}) and using the definition of $\tau_{\eta}$
given in Eq.~(B5) of Ref.~\cite{LK}, we find that the collision term reduces to
\begin{equation}
\left\langle
\frac{1}{m}\left(p_{\mu}p_{\nu}-\frac{1}{3}\delta_{\mu\nu}p^2
\right)
\right\rangle_{\rm coll}=
-\frac{P_{\mu\nu}'}{\tau_{\eta}},
\end{equation}
where the viscous relaxation time $\tau_{\eta}$ is defined in Refs.~\cite{LK,relax}.
Assuming the harmonic time dependence $P_{\mu\nu}'\propto e^{-i\omega t}$,
we finally obtain
\begin{equation}
P_{\mu\nu}'=-\frac{2\tau_{\eta}\tilde P_0}{1-i\omega\tau_{\eta}}
\left(D_{\mu\nu}-\frac{1}{3}\delta_{\mu\nu}{\rm Tr} D \right)
=-2\eta(\omega)\left(D_{\mu\nu}-\frac{1}{3}\delta_{\mu\nu} {\rm Tr} D
\right),
\end{equation}
where the frequency-dependent viscosity coefficient $\eta(\omega)$ is defined by
\begin{equation}
\eta(\omega)\equiv\frac{\tau_{\eta}\tilde P_0}{1-i\omega\tau_{\eta}}
=\frac{\eta}{1-i\omega\tau_{\eta}}.
\end{equation}

The frequency-dependent thermal conductivity can also be obtain in the same manner,
by considering the linearized heat current
\begin{eqnarray}
{\bf Q}({\bf r},t)&=&\int\frac{d{\bf p}}{(2\pi\hbar)^3}
\frac{p^2}{2m}\frac{{\bf p}}{m}
f({\bf r},{\bf p},t)-\frac{5}{2}{\bf v}_n\tilde P_0({\bf r}) \cr
&=&\int\frac{d{\bf p}}{(2\pi\hbar)^3}
\left[\frac{p^2}{2m}-\frac{5}{2}k_{\rm B}T_0\frac{g_{5/2}(z_0)}{g_{3/2}(z_0)}\right]
\frac{{\bf p}}{m}f({\bf r},{\bf p},t).
\end{eqnarray}
Taking the moment of the kinetic equation in Eq.~(\ref{kineq_leq}),
one finds that the relevant contribution is given by
\begin{eqnarray}
&&\int \frac{d{\bf p}}{(2\pi\hbar)^3}\left[\frac{p^2}{2m}
-\frac{5}{2}k_{\rm B}T_0\frac{g_{5/2}(z_0)}{g_{3/2}(z_0)}\right]
\frac{{\bf p}}{m}\frac{p^2}{2mk_{\rm B}T_0^2}
\left(\frac{{\bf p}}{m}\cdot\bm{\nabla}\delta T\right)f_0(1+f_0) \cr
&&=\frac{5}{2}\frac{\tilde n_0k_{\rm B}^2T_0}{m}
\bm{\nabla}\delta T\left\{\frac{7g_{7/2}(z_0)}{2g_{3/2}(z_0)}
-\frac{5}{2}\left[\frac{g_{5/2}(z_0)}{g_{3/2}(z_0)}\right]^2\right\}.
\end{eqnarray}
One thus obtains
\begin{equation}
\frac{\partial {\bf Q}}{\partial t}+
\frac{5}{2}\frac{\tilde n_0k_{\rm B}^2T_0}{m}
\bm{\nabla}\delta T\left\{\frac{7g_{7/2}(z_0)}{2g_{3/2}(z_0)}
-\frac{5}{2}\left[\frac{g_{5/2}(z_0)}{g_{3/2}(z_0)}\right]^2\right\}=
\left\langle\left[\frac{p^2}{2m}-\frac{5}{2}k_{\rm B}T_0
\frac{g_{5/2}(z_0)}{g_{3/2}(z_0)}\right]\frac{{\bf p}}{m} \right\rangle_{\rm coll}.
\end{equation}
where
\begin{equation}
\left\langle
\left[\frac{p^2}{2m}-\frac{5}{2}k_{\rm B}T_0
\frac{g_{5/2}(z_0)}{g_{3/2}(z_0)}\right]
\frac{{\bf p}}{m} \right\rangle_{\rm coll}
\equiv
\int \frac{d{\bf p}}{(2\pi\hbar)^3}
\left[\frac{p^2}{2m}-\frac{5}{2}k_{\rm B}T_0
\frac{g_{5/2}(z_0)}{g_{3/2}(z_0)}\right]\frac{{\bf p}}{m}(C_{12}+C_{22}).
\label{q_coll}
\end{equation}
To evaluate the collisional term Eq.~(\ref{q_coll}), which arises from deviation from local equilibrium,
we use the Chapman-Enskog ansatz $f=f_{\rm leq}+\delta f$, where \cite{LK}
\begin{equation}
\delta f = {\bf A}\cdot\frac{{\bf p}}{m}
\left[\frac{p^2}{2m}-\frac{5}{2}k_{\rm B}T_0
\frac{g_{5/2}(z_0)}{g_{3/2}(z_0)}\right]f_0(1+f_0).
\label{ansatz_q}
\end{equation}
Here ${\bf A}$ is a momentum-independent vector which is directly related to the heat current
${\bf Q}$ through
\begin{equation}
{\bf Q}=
{\bf A}\int \frac{d{\bf p}}{(2\pi\hbar)^3}\frac{p^2}{3m}
\left[\frac{p^2}{2m}-\frac{5}{2}k_{\rm B}T_0
\frac{g_{5/2}(z_0)}{g_{3/2}(z_0)}\right]^2f_0(1+f_0).
\label{QandA}
\end{equation}
Evaluation of the collisional term with using the ansatz Eq.~(\ref{ansatz_q}) closely
follows the derivation of the thermal conductivity in Ref.~\cite{LK}.
We find
\begin{eqnarray}
&&\left\langle\left[\frac{p^2}{2m}-\frac{5}{2}k_{\rm B}T_0
\frac{g_{5/2}(z_0)}{g_{3/2}(z_0)}\right]\frac{{\bf p}}{m} \right\rangle_{\rm coll} \cr&
&=\frac{{\bf A}}{3}\int 
\left[
\frac{p^2}{2m}-\frac{5}{2}k_{\rm B}T_0
\frac{g_{5/2}(z_0)}{g_{3/2}(z_0)}
\right]\frac{{\bf p}}{m}
\cdot\hat{L}\left[\left\{
\frac{p^2}{2m}-\frac{5}{2}k_{\rm B}T_0
\frac{g_{5/2}(z_0)}{g_{3/2}(z_0)} \right\}\frac{{\bf p}}{m}\right].
\label{QcollandA}
\end{eqnarray}
Combining Eq.~(\ref{QandA}) and Eq.~(\ref{QcollandA}) and using the definition of the thermal
relaxation time $\tau_{\kappa}$ given in Eq.~(B1) of Ref.~\cite{LK}, we find
\begin{equation}
\left\langle\left[\frac{p^2}{2m}-\frac{5}{2}k_{\rm B}T_0
\frac{g_{5/2}(z_0)}{g_{3/2}(z_0)}\right]\frac{{\bf p}}{m} \right\rangle_{\rm coll}
=-\frac{{\bf Q}}{\tau_{\kappa}}.
\end{equation}
Assuming the harmonic time dependence ${\bf Q}\propto e^{-\omega t}$, we obtain
\begin{equation}
{\bf Q}=-\frac{\kappa}{1-i\omega\tau_{\kappa}}\bm{\nabla}\delta T
\equiv -\kappa(\omega)\bm{\nabla}\delta T,
\end{equation}
where we have used the expression for $\kappa$ given in Eq.~(\ref{eq:kappa}),
and the frequency-dependent thermal conductivity is defined by
\begin{equation}
\kappa(\omega)\equiv \frac{\kappa}{1-i\omega\tau_{\kappa}}.
\end{equation}

\section{moment method for a degenerate normal Bose gas}

\begin{figure}
\centerline{\epsfig{file=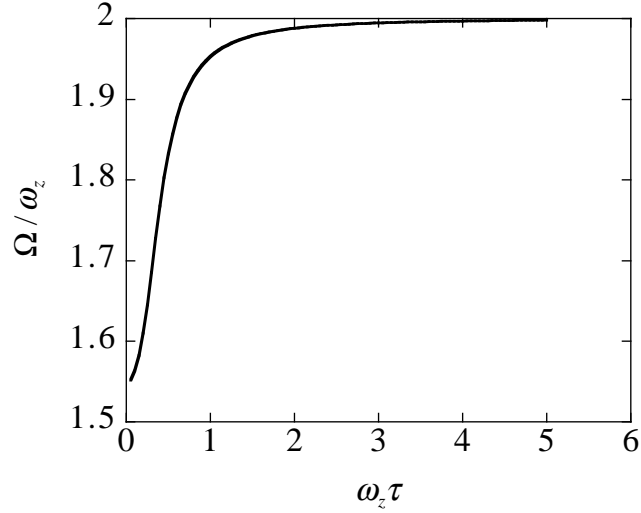,height=3.0in}}
\centerline{\epsfig{file=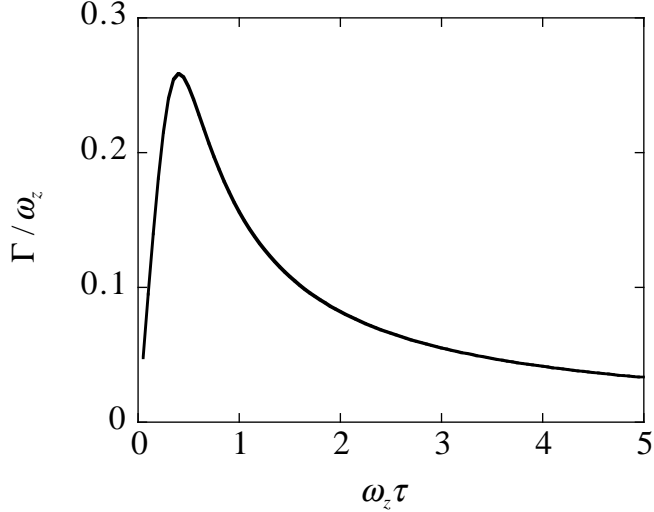,height=3.0in}}
\caption{Moment calculation of the (a) frequency 
and (b) damping of the $m=0$ mode, both as a function of the quadrupole relaxation time $\tau$.
These results are found by solving Eq.~(\ref{freq_moment}).
We use the ENS trap frequencies $\omega_{\perp}/2\pi=988$Hz and $\omega_z/2\pi=115$Hz
\cite{ENS1,ENS2,ENS3}.}.
\label{moment}
\end{figure}

\begin{figure}
\centerline{\epsfig{file=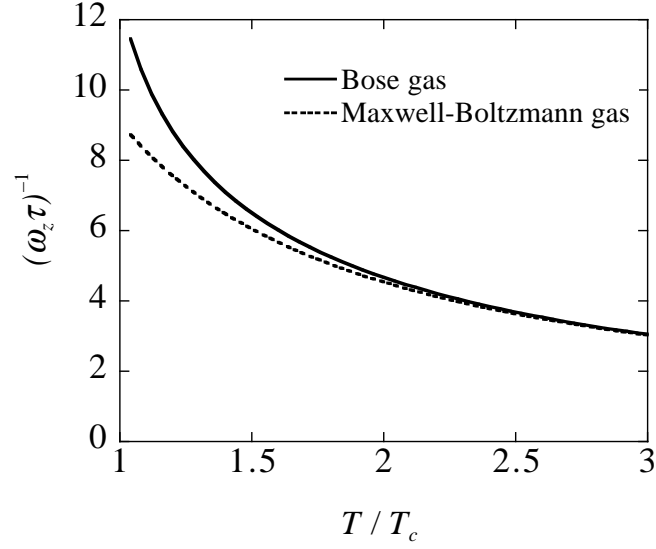,height=3.0in}}
\caption{Temperature dependence of the spatially-averaged quadrupole relaxation time $\tau$.
The broken line shows the result using the Maxwell-Boltzmann distribution for a classical
trapped gas.}
\label{moment2}
\end{figure}

\begin{figure}
\centerline{\epsfig{file=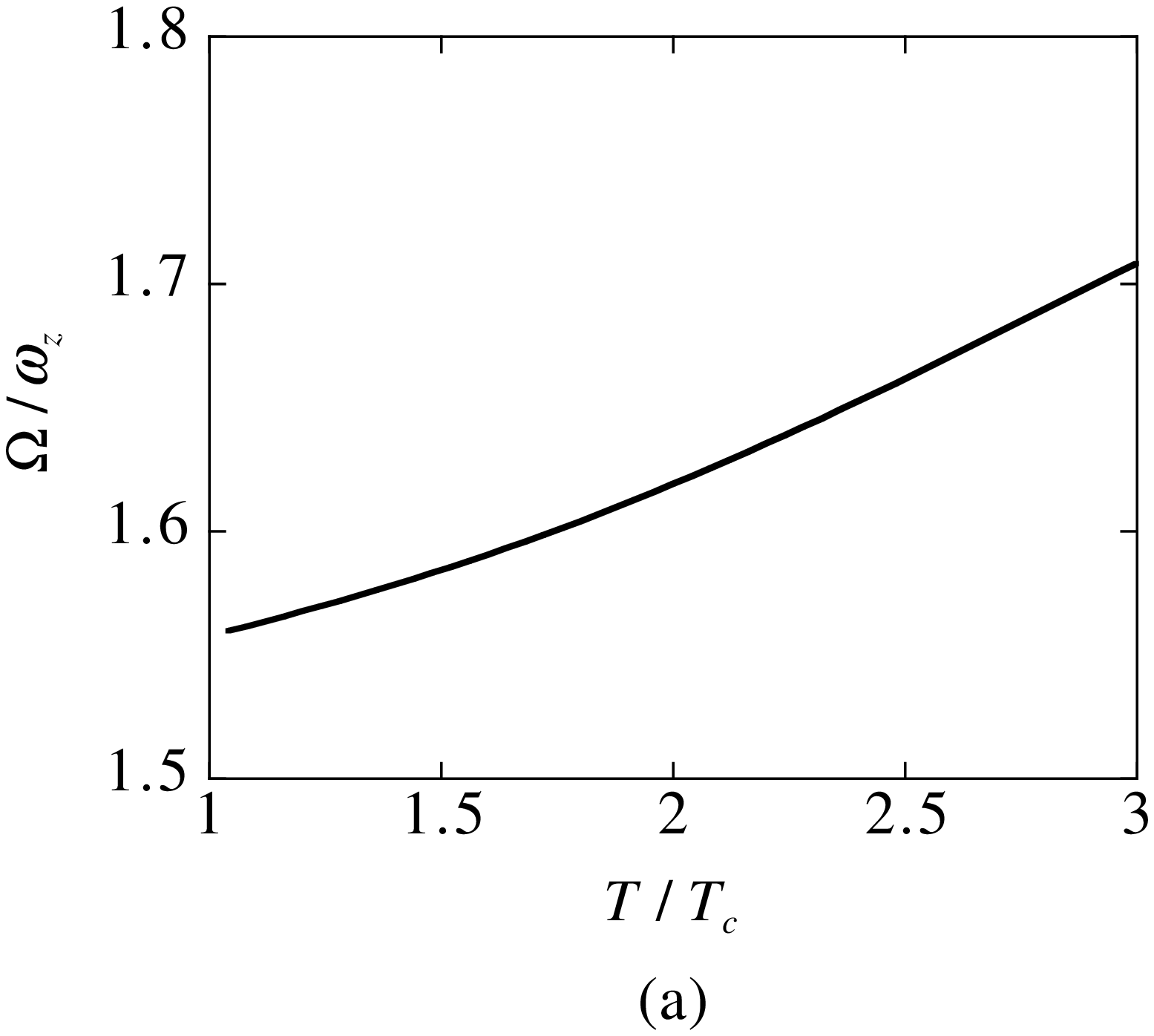,height=3.0in}}
\centerline{\epsfig{file=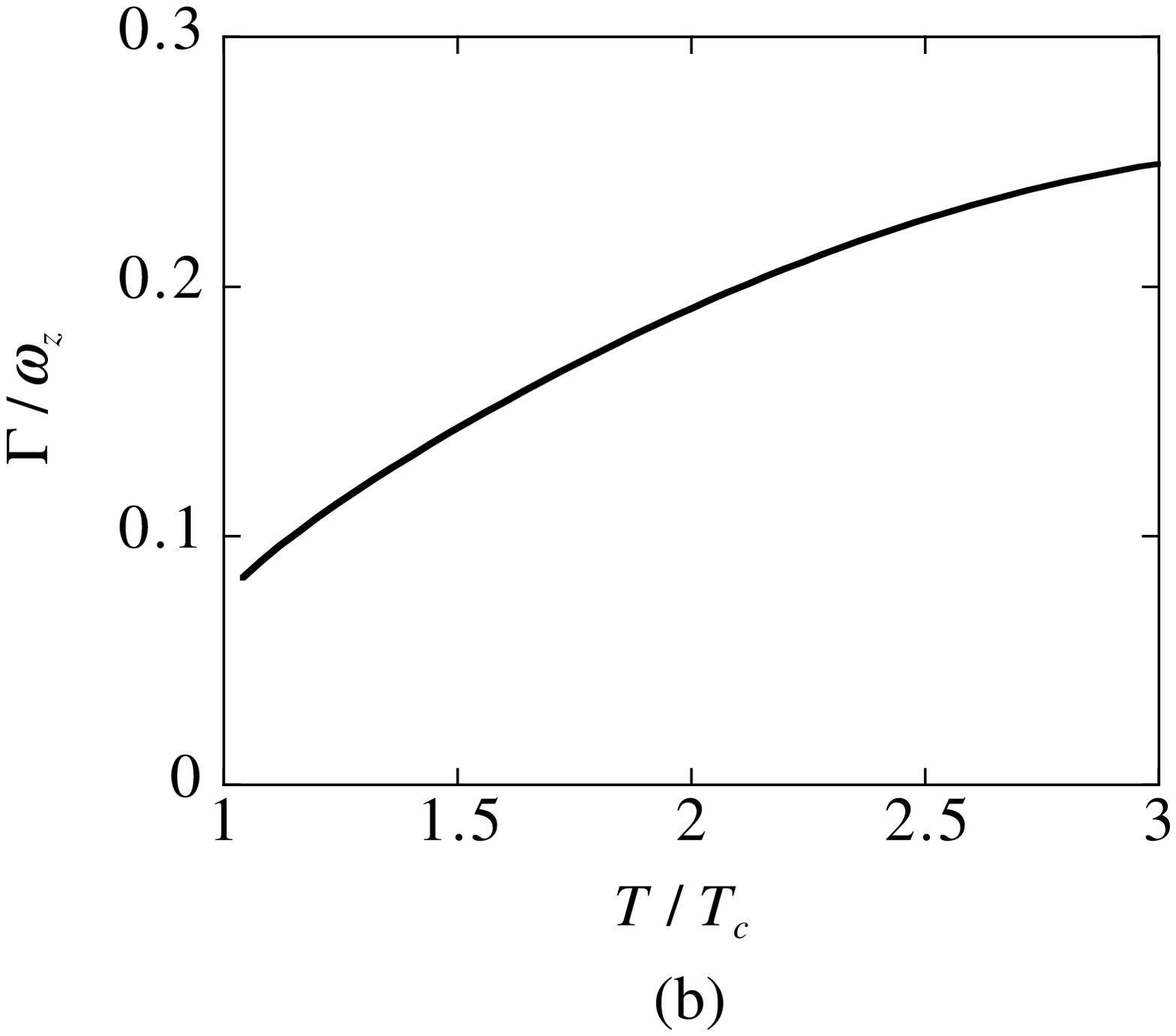,height=3.0in}}
\caption{Temperature dependence of (a) frequency $\Omega$, and (b) damping $\Gamma$
of the $m=0$ mode, obtained by the moment method.}
\label{moment3}
\end{figure}

In the moment method \cite{moment}, one derives an equation of motion for some dynamical quantity
denoted by $\chi ({\bf r},t)$ by taking a moment of the kinetic equation in both position
and momentum:
\begin{equation}
\langle\chi\rangle\equiv\frac{1}{N}\int d{\bf r}\frac{d{\bf p}}{(2\pi\hbar)^3}
\chi({\bf r},t)f({\bf r},{\bf p},t).
\end{equation}
The advantage of this method is that it gives results valid in both the collisionless and
hydrodynamic region.
It should be noted that in the collisionless region, this approach does not include Landau
damping, but this is small above $T_{\rm BEC}$.
For the $m=0$ monopole-quadrupole mode, we need moment equations for the following
physical quantities:
\begin{eqnarray}
&&\chi_1=r^2,~~\chi_2=2z^2-r^2, \cr
&&\chi_3={\bf r}\cdot{\bf p}/m,~~\chi_4=2zp_z/m-{\bf r}_{\perp}\cdot{\bf p}_{\perp}/m, \cr
&&\chi_5=p^2/m^2,~~\chi_6=2p_z^2/m^2-p_{\perp}^2/m^2.
\end{eqnarray}
Calculation of these moments give the following coupled equations
\begin{eqnarray}
&&\frac{d\langle\chi_1\rangle}{dt}-2\langle\chi_3\rangle=0,~~
\frac{d\langle\chi_2\rangle}{dt}-2\langle\chi_4\rangle=0,\cr
&&\frac{d\langle\chi_3\rangle}{dt}-\langle \chi_5\rangle+
\frac{2\omega_{\perp}^2+\omega_z^2}{3}\langle\chi_1\rangle
+\frac{\omega_z^2-\omega_{\perp}^2}{3}\langle\chi_2\rangle=0,\cr
&&\frac{d\langle\chi_4\rangle}{dt}-\langle \chi_6\rangle+
\frac{2\omega_z^2-2\omega_{\perp}^2}{3}\langle\chi_1\rangle
+\frac{\omega_{\perp}^2+2\omega_z^2}{3}\langle\chi_2\rangle=0,\cr
&&\frac{d\langle\chi_5\rangle}{dt}+\frac{2\omega_z^2+4\omega_{\perp}^2}{3}
\langle\chi_3\rangle+\frac{2\omega_z^2-\omega_{\perp}^2}{3}
\langle\chi_4\rangle=0, \cr
&&\frac{d\langle\chi_6\rangle}{dt}+\frac{4\omega_z^2-4\omega_{\perp}^2}{3}
\langle\chi_3\rangle+\frac{4\omega_z^2+2\omega_{\perp}^2}{3}
\langle\chi_4\rangle=\langle\chi_6\rangle_{\rm coll}. 
\label{moment_eqs}
\end{eqnarray}

The collisional contribution in the last equation in (\ref{moment_eqs}) is defined as
\begin{equation}
\langle \chi_6 \rangle_{\rm coll}\equiv \frac{1}{N}\int d{\bf r}
\int \frac{d{\bf p}}{(2\pi\hbar)^3}\chi_6 C_{22}[f].
\label{chi6coll}
\end{equation} 
This term can be approximately evaluated by using the ansatz $f=f_0+\delta f$, where
\begin{equation}
\delta f=\beta_0f_0(1+f_0)
\left[\frac{\delta T}{T_0}\left(\frac{p^2}{2m}+U_0-\mu_{c0}\right)
+{\bf p}\cdot{\bf v}_n+\delta\tilde\mu
+\alpha(2p_z^2-p^2)\right],
\label{moment_ansatz}
\end{equation}
where the time-dependent parameter $\alpha(t)$ characterizes the anisotropy
in the momentum distribution described by $\langle\chi_6\rangle$.
The above ansatz is a generalization of the gaussian ansatz for the Maxwell-Boltzmann
gas used in Ref.~\cite{moment} to a degenerate Bose gas.

Using the ansatz in Eq.~(\ref{moment_ansatz}) in Eq.~(\ref{chi6coll}) and linearizing in $\alpha$,
one obtains
\begin{equation}
\langle\chi_6\rangle_{\rm coll}=-\frac{\langle\chi_6\rangle}{\tau},
\label{eq:chi6}
\end{equation}
where $\tau$ is a quadrupole relaxation time defined by the weighted spatial
average of the inverse of the viscous relaxation time (see also Ref.~\cite{Nikuni})
\begin{equation}
\frac{1}{\tau}\equiv \frac{\int d{\bf r} \tilde P_0/\tau_{\eta}}
{\int d{\bf r}\tilde P_0}.
\label{tau_q}
\end{equation}
Using the moment equations Eq.~(\ref{moment_eqs}) together with the approximation Eq.~(\ref{eq:chi6}),
the only effect of using Bose statistics is in the value for the spatially-averaged 
relaxation time $\tau$ in Eq.~(\ref{tau_q}).

In a nondegenerate (Maxwell-Boltzmann) gas, one finds that
$\tau_{\eta}({\bf r})=\frac{5}{4}\tau_{\rm cl}({\bf r})$ \cite{LK,relax,SJ},
where $\tau_{\rm cl}$ is the usual elastic collision time for a classical gas
\begin{equation}
\tau_{\rm cl}^{-1}({\bf r})=\sqrt{2}\sigma\tilde n_0({\bf r})\bar v=
\sqrt{2}(8\pi a^2)\tilde n_0({\bf r})
(8k_{\rm B}T/\pi m)^{1/2}.
\end{equation}
Using this result and $\tilde P_0({\bf r})=k_{\rm B}T\tilde n_0({\bf r})$ in Eq.~(\ref{tau_q}) ,
the quadrupole relaxation time reduces to
\begin{equation}
\frac{1}{\tau}=\frac{4}{5}\Gamma_{\rm coll}.
\label{taucl}
\end{equation}
Here $\Gamma_{\rm coll}\equiv \sqrt{2}\sigma\bar n\bar v$, where the spatially-averaged
density is defined by
\begin{equation}
\bar n\equiv\frac{\displaystyle \int d{\bf r}\tilde n_0^2({\bf r})}
{\displaystyle \int d{\bf r}\tilde n_0({\bf r})}
=\frac{\tilde n_0({\bf r}={\bf 0})}{2\sqrt{2}}.
\end{equation}
This result for $1/\tau$ in terms of $\Gamma_{\rm coll}$ agrees with that of
Gu\'ery-Odelin et al.~\cite{moment}.
The authors of Ref.~\cite{ENS3} analyzed data in terms of the classical gas result of Ref.~\cite{moment},
using the above definition of the averaged collision rate $\Gamma_{\rm coll}$ valid for a
Maxwell-Boltzmann gas.

Assuming the time dependence $e^{-i\omega t}$, the coupled equations in Eq.~(\ref{moment_eqs})
with Eq.~(\ref{eq:chi6}) can be solved to give
\begin{equation}
(\omega^2-4\omega_z^2)(\omega^2-4\omega_{\perp}^2)+\frac{i}{\omega\tau}
\left[\omega^4-\frac{2}{3}\omega^2(5\omega^2_{\perp}+4\omega_z^2)
+8\omega_{\perp}^2\omega_z^2\right]=0.
\label{freq_moment}
\end{equation}
Solving Eq.~(\ref{freq_moment}), we obtain the solution $\omega=\Omega-i\Gamma$,
describing damped modes.
One can see that in the collisionless limit $\omega\tau\gg 1$, the frequency
is given by either $2\omega_z$ or $2\omega_{\perp}$.
In the opposite hydrodynamic limit $\omega\tau\ll 1$, the two solutions are given by
$\omega=\Omega_{\pm}$, with $\Omega_{\pm}$ given by Eq.~(\ref{freq_hydro}).
The frequency $\Omega$ and damping $\Gamma$ obtained by
solving Eq.~(\ref{freq_moment}) can be expressed in terms of the spatially-averaged quadrupole
relaxation time $\tau$, for given trap frequencies.
In Fig.~\ref{moment}, we plot $\Omega$ and $\Gamma$ of the low-frequency mode as a function of 
$\omega_z\tau$.
The hydrodynamic domain is the region $\omega_z\tau \lesssim 1$.

In Fig.~\ref{moment2}, we plot the temperature dependence of the
quadrupole relaxation time $\tau$ calculated for the ENS experimental data.
For comparison, we also plot the result \cite{ENS3,moment}using the Maxwell-Boltzmann distribution.
The effect of Bose statistics is clearly very small, down to about $T\simeq 1.5T_c$.
We use the result in Fig.~\ref{moment2} to calculate the frequency and damping as functions
of the temperature, as shown in Fig.~\ref{moment3}.

In the hydrodynamic limit, one can obtain an analytical expression for the damping rate
from the moment equations.
Assuming $\omega\tau\ll1$, one can expand the solution to
Eq.~(\ref{freq_moment}) to first order in $\tau$.
In this limit, the damping $\Gamma_-$ of the low-frequency mode $\Omega_-$ is given by
\begin{equation}
\Gamma_-=\frac{\tau}{2(\Omega^2_+-\Omega^2_-)}
(\Omega_-^2-4\omega_z^2)(\Omega_-^2-4\omega^2_{\perp}).
\label{gamma_moment}
\end{equation}
Apart from a different averaged shear-viscous relaxation time $\tau$, it is satisfying that
this moment result is identical to the LL expression in Eq.(\ref{gamma_hydro}).

\end{appendix}

\noindent

\end{document}